\documentclass[prb,aps,twocolumn,superscriptaddress,10pt]{revtex4-2}
\usepackage[utf8]{inputenc}
\usepackage[T1]{fontenc}
\usepackage{color}
\usepackage{amsmath}
\usepackage{amsfonts}
\usepackage{xr-hyper}
\usepackage{hyperref}
\usepackage{amsthm}
\usepackage{bm}
\usepackage{float}
\usepackage{textcomp}
\usepackage{upgreek}
\usepackage{textgreek}
\usepackage{setspace}
\usepackage[percent]{overpic}
\usepackage[table]{xcolor}

\hypersetup{
  colorlinks = true,
  allcolors= blue,
}
\usepackage{tikz,pgfplots}
\usepackage{blkarray}
\usetikzlibrary{patterns}
\usetikzlibrary{fadings}
\usetikzlibrary{decorations.pathmorphing,patterns}
\tikzset{snake it/.style={decorate, decoration=snake}}
\usetikzlibrary{arrows, decorations.markings}
\pgfplotsset{compat=1.14}
\tikzset{
vecArrow/.style={
  thick,
  decoration={markings,mark=at position
   1 with {\arrow[scale=2,thin]{open triangle 60}}},
  double distance=1.4pt, shorten >= 10.5pt,
  preaction = {decorate},
  postaction = {draw,line width=1.4pt, white,shorten >= 4.5pt}
  },
innerWhite/.style={
  semithick,
  white,
  line width=1.4pt,
  shorten >= 4.5pt
  }
}

\definecolor{orange}{rgb}{1,0.5,0}
\definecolor{darkgreen}{rgb}{0,0.4,0.1}
\definecolor{cola}{HTML}{5FBDFF}
\definecolor{colb}{HTML}{118DFE}
\definecolor{cold}{HTML}{EEFF0C}
\definecolor{cole}{HTML}{FC9E0A}

\newcommand{\WidthFigure}{\columnwidth}
\newcommand{\folder}{./}

\makeatletter
\newcommand{\doublehat}[1]{%
\begingroup%
  \let\macc@kerna\z@%
  \let\macc@kernb\z@%
  \let\macc@nucleus\@empty%
  \hat{\raisebox{.3ex}{\vphantom{\ensuremath{#1}}}\smash{\hat{#1}}}%
\endgroup%
}
\makeatother

\makeatletter
\newcommand{\doublehatSub}[1]{%
\begingroup%
  \let\macc@kerna\z@%
  \let\macc@kernb\z@%
  \hat{\raisebox{-.07ex}{\vphantom{\ensuremath{#1}}}\smash{\hat{#1}}}%
\endgroup%
}
\makeatother

\makeatletter
\DeclareFontFamily{OMX}{MnSymbolE}{}
\DeclareSymbolFont{MnLargeSymbols}{OMX}{MnSymbolE}{m}{n}
\SetSymbolFont{MnLargeSymbols}{bold}{OMX}{MnSymbolE}{b}{n}
\DeclareFontShape{OMX}{MnSymbolE}{m}{n}{
    <-6>  MnSymbolE5
   <6-7>  MnSymbolE6
   <7-8>  MnSymbolE7
   <8-9>  MnSymbolE8
   <9-10> MnSymbolE9
  <10-12> MnSymbolE10
  <12->   MnSymbolE12
}{}
\DeclareFontShape{OMX}{MnSymbolE}{b}{n}{
    <-6>  MnSymbolE-Bold5
   <6-7>  MnSymbolE-Bold6
   <7-8>  MnSymbolE-Bold7
   <8-9>  MnSymbolE-Bold8
   <9-10> MnSymbolE-Bold9
  <10-12> MnSymbolE-Bold10
  <12->   MnSymbolE-Bold12
}{}

\let\llangle\@undefined
\let\rrangle\@undefined
\DeclareMathDelimiter{\llangle}{\mathopen}%
                     {MnLargeSymbols}{'164}{MnLargeSymbols}{'164}
\DeclareMathDelimiter{\rrangle}{\mathclose}%
                     {MnLargeSymbols}{'171}{MnLargeSymbols}{'171}
\makeatother
\DeclareUnicodeCharacter{300}{à}

\DeclareMathAlphabet{\mathsfit}{\encodingdefault}{\sfdefault}{m}{sl}
\SetMathAlphabet{\mathsfit}{bold}{\encodingdefault}{\sfdefault}{bx}{sl}
\newcommand{\tens}[1]{\bm{\mathsfit{#1}}}
\newcommand{\tenscomp}[1]{\mathsfit{#1}}

\makeatletter
\usepackage{comment}
\let\wfs@comment@comment\comment
\let\comment\@undefined

\usepackage[final]{changes}
\let\wfs@changes@comment\comment
\let\comment\@undefined

\newcommand\comment{%
    \ifthenelse{\equal{\@currenvir}{comment}}
    {\wfs@comment@comment}
    {\wfs@changes@comment}%
}
\makeatother

\definecolor{dgreen}{rgb}{0,0.45,0}

\makeatletter
\@namedef{Changes@AuthorColor}{red}
\colorlet{Changes@Color}{red}
\setaddedmarkup{\textcolor{orange}{#1}}
\makeatother

\begin{document}

\title{
Vibrational and thermal properties of amorphous alumina from first principles
}

\author{Angela F. Harper}
\altaffiliation{Equally contributed}
\altaffiliation{Current address: Fritz-Haber Institut der Max Planck Gesellschaft, Berlin (DE)}
\affiliation{Theory of Condensed Matter Group, Cavendish Laboratory, University of Cambridge (UK)}

\author{Kamil Iwanowski}
\altaffiliation{Equally contributed}
\affiliation{Theory of Condensed Matter Group, Cavendish Laboratory, University of Cambridge (UK)}

\author{William C. Witt}
\affiliation{Department of Materials Science \& Metallurgy, University of Cambridge, Cambridge,
(UK)}

\author{Mike C. Payne}
\affiliation{Theory of Condensed Matter Group, Cavendish Laboratory, University of Cambridge (UK)}

\author{Michele Simoncelli}
\email{ms2855@cam.ac.uk}
\affiliation{Theory of Condensed Matter Group, Cavendish Laboratory, University of Cambridge (UK)}

\begin{abstract}
Amorphous alumina is employed ubiquitously as a high-dielectric-constant material in electronics, and
its thermal-transport properties are of key relevance for heat management in electronic chips and devices. Experiments show that the thermal conductivity of alumina depends significantly on the synthesis process, indicating the need for a theoretical study to elucidate the atomistic origin of these variations. 
Here we employ first-principles simulations to characterize the atomistic structure, vibrational properties, and thermal conductivity of alumina at densities ranging from 2.28 g/cm$^3$ to 3.49 g/cm$^3$. 
Moreover, using an interatomic potential trained on first-principles data, we investigate how system size affects predictions of the thermal conductivity, showing that simulations containing 120 atoms can already reproduce the bulk limit of the conductivity.  
Finally, relying on the recently developed Wigner formulation of thermal transport, we shed light on  
the interplay between atomistic topological disorder and anharmonicity in the context of heat conduction, showing that the former dominates over the latter in determining the conductivity of  alumina. 
\end{abstract}
\maketitle

\section{Introduction} 
\label{sec:intro}

Alumina (Al$_2$O$_3$) is a deceptively complex material; there are at least nine known metastable polymorphs \cite{levin_1998_polymorphs}, an amorphous phase which possesses its own unique properties as a result of local disorder \cite{tavakoli2013amorphous}, and a number of predicted non-stoichiometric structures \cite{li_effects_2020}. 
Alumina has applications in catalysis \cite{shafiq2022recent,oh2019sustainable}, energy-storage devices (e.g., Li-ion batteries \cite{jin2019li4ti5o12}, Li-metal anodes \cite{qu2019air}, and solid state electrolytes \cite{randau2021additive}), as well as in heat-management technologies \cite{paterson_thermal_2020} and coatings \cite{mavric_advanced_2019}. 
Moreover, thin film amorphous alumina (am-Al$_2$O$_3$) is often employed in electronic devices, and its thermal properties play a crucial role in determining both device efficiency and lifespan \cite{paterson_thermal_2020,scott2018thermalal2o3}.
The structural properties of am-Al$_2$O$_3$ have become subject of extensive research, since the atomic-layer deposition (ALD) technique allows production of alumina films with controlled thickness and morphology.
There are several open fundamental questions on how the structure of am-Al$_2$O$_3$ affects its macroscopic properties. Solid-state $^{27}$Al nuclear magnetic resonance has identified three main aluminum coordination environments in am-Al$_2$O$_3$ \cite{fharper_modelling_2023,Lee_2010_NMR,sarou_2013_thermal,hashimoto_structure_2022,shi2019structure}, which play a role in the electronic transport as evidenced by X-Ray absorption spectroscopy and first principles electronic density of states calculations \cite{Dicks_2019,Leung_2021}. 
The literature on the relationships between structural, vibrational, and thermal properties of am-Al$_2$O$_3$ at technologically relevant temperatures (\textit{i.e.} at $T> 30$ K, in the so called above-the-plateau regime \cite{cahill_thermal_1987,allen1989thermal,simoncelli_thermal_2022}) is less developed.
Recently, \citet{li_effects_2020} used machine learned force-fields with molecular dynamics (Green-Kubo (GKMD \cite{ercole_gauge_2016,knoop_ab_2023,knoop_anharmonicity_2023,marcolongo2016microscopic,eriksson_tuning_2023,lv2016_locons,donadio2009atomistic}) and non-equilibrium (NEMD \cite{rhahn_thermal_2022,Jund1999,McGaughey2009predicting,felix_thermal_2018}) approaches) to compute the room-temperature thermal conductivity of atomistic models of am-AlO$_x$; they studied models with densities ranging from 2.6 to 3.3 g/cm$^3$ and containing up to 528 atoms.
While providing insights on the physics governing thermal transport at room temperature, this work could not explore how the thermal conductivity varies with temperature---in fact, GKMD and NEMD are both governed by classical equipartition and thus cannot be used to investigate how the conductivity changes as temperature decreases (\textit{i.e.}, when the specific heat deviates significantly from the classical limit \cite{PhysRevMaterials.3.085401}). 

Here we employ first-principles calculations alongside a MACE machine learning potential (MLP) \cite{batatia2022mace, kovacs2023evaluation} to characterize the structural, vibrational, and thermal properties of am-Al$_2$O$_3$ across a wide range of densities ($2.28{\leq}\rho{\leq}3.49$ g/cm$^3$). (The MACE architecture utilizes the Atomic Cluster Expansion  \cite{drautz_atomic_2019, dusson2022atomic, witt_acepotentialsjl_2023} and equivariant message passing.) In particular, in Sec.~\ref{sec:structural_properties_of_amor} we investigate from first principles how the atomic coordination topology of am-Al$_2$O$_3$ changes with density, and how these changes affect the vibrational properties.
The vibrational properties are then used in Sec.~\ref{sec:thermal_properties} as inputs for the Wigner formulation of thermal transport \cite{simoncelli2019unified,simoncelli2021Wigner}, which can be used to calculate heat conduction in amorphous solids accounting comprehensively for the effects of structural disorder, anharmonicity, and quantum Bose-Einstein statistics of vibrations \cite{simoncelli_thermal_2022}. We employ the convergence-acceleration protocol discussed in Ref.~\cite{simoncelli_thermal_2022} to compute the bulk limit of the thermal conductivity of am-Al$_2$O$_3$, highlighting the agreement between our predictions and experiments  at various densities and temperatures.
We describe how the interplay between structural disorder and anharmonicity affects the thermal conductivity of am-Al$_2$O$_3$, demonstrating that disorder practically determines thermal transport at all the densities and temperatures studied ($50{\leq}T{\leq}700$ K).
Finally, in Sec. \ref{sec:size_effects} we employ the MACE MLP to study size effects, 
comparing the structural, vibrational, and thermal properties of models of am-Al$_2$O$_3$ containing 120, 3240, and 7680 atoms. We discuss how the convergence-acceleration protocol of Ref.~\cite{simoncelli_thermal_2022} affects the thermal-conductivity predictions in models having different sizes.

\begin{figure}[b]
  \centering
\includegraphics[width=\WidthFigure]{\folder/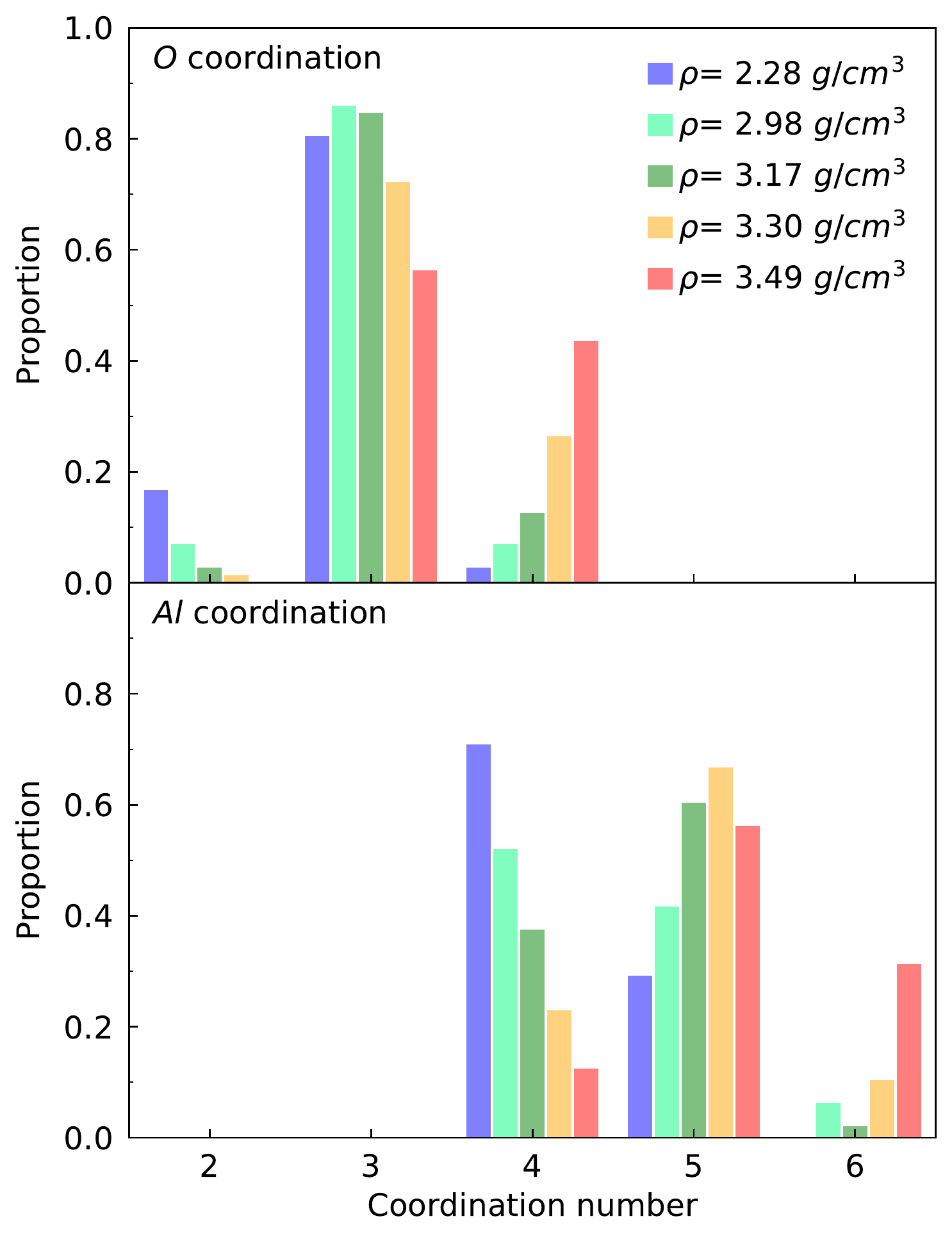}\vspace*{-3mm}
  \caption{\textbf{Coordination of oxygen and aluminum in am-Al$_2$O$_3$ at various densities.} 
  The histograms show the probability of oxygen (top) or aluminum (bottom) to have a certain coordination number in the disordered models analyzed (different densities are distinguished with colors).
  Coordination varies from 2 to 4 for O, and from 4 to 6 for Al.}
  \label{fig:coord}
\end{figure}
\section{Structural and vibrational properties}
\label{sec:structural_properties_of_amor}

The properties of am-Al$_2$O$_3$ strongly depend on the synthesis process---samples with densities ranging from 2.1 g/cm$^3$ to 3.6 g/cm$^3$ have been synthesized and discussed in the literature~\cite{bhatia1989alumina,koski_properties_1999}. To generate structures throughout this range, we extracted 120-atom structures at 3.17 g/cm$^3$, 3.30 g/cm$^{3}$ and 3.49 g/cm$^{3}$ from Ref. \cite{fharper_modelling_2023}, and additionally generated two 120-atom am-Al$_2$O$_3$ models at 2.28 g/cm$^3$ and 2.98 g/cm$^{3}$ using a melt-quench procedure~\cite{fharper_modelling_2023}. Each structure was generated using \textit{ab initio} molecular dynamics (AIMD): initially melted at a temperature of 4000\,K, then quenched to a temperature of 300\,K, and finally equilibrated in the NVT ensemble (see Appendix~\ref{sec:computational_details} for details). 
In the following, we show that these structures have very different local atomic environments and coordination topologies, and consequently different vibrational and thermal properties.

\subsection{Coordination Topology} 
\label{sub:coordination_topology}

With the goal of gaining insights on how the atomic bonding topology affects the macroscopic conductivity,
we start by characterizing the coordination environments present in our am-Al$_2$O$_3$ models, showing in Fig.~\ref{fig:coord} the probability distribution of finding $x$-fold-coordinated oxygen ($\rm O_x$) or $y$-fold-coordinated aluminum ($\rm Al_y$) at each density.
Six different coordination environments are present in our models---$\rm O_x$ with $x\in 2,3,4$, and $\rm Al_y$ with $y\in 4,5,6$---and at least five out of these six coordination topologies coexist at every model density analyzed.

\begin{figure}[b]
  \centering
\includegraphics[width=\WidthFigure]{\folder/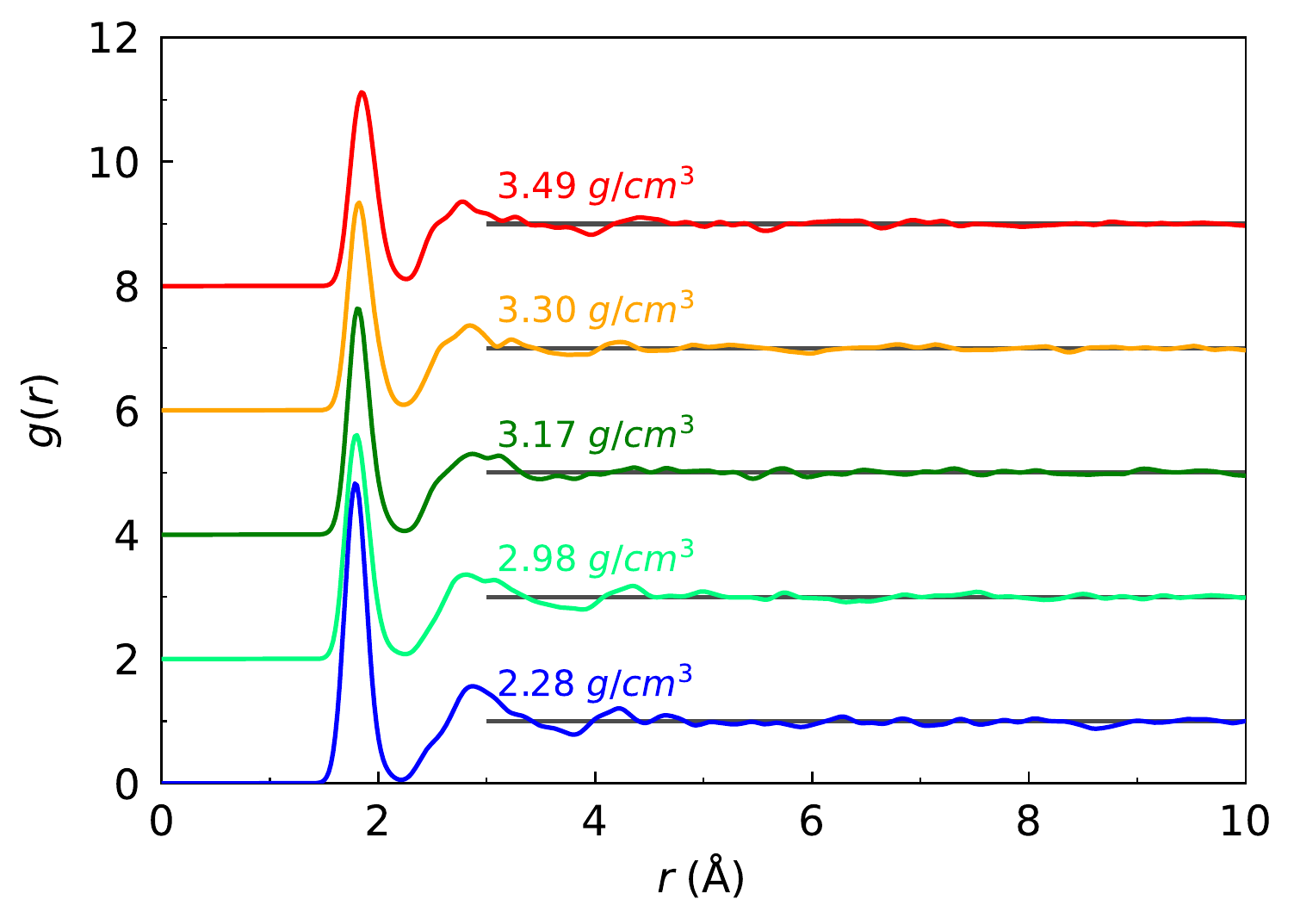}\\[-3mm]
  \caption{
  \textbf{Radial distribution function at various densities.}  Solid lines, total RDF of the am-Al$_2$O$_3$ generated and studied in this work (a rigid shift of two is used to distinguish the RDF of different models).
  }
  \label{fig:al2o3_rdf}
\end{figure}

As shown in the top panel of Fig. \ref{fig:coord}, at increasing densities
the concentration of twofold-coordinated oxygens (O$_2$) generally decreases in favor of fourfold-coordinated (O$_4$);
changes in density have weaker effects on the concentration of threefold-coordinated oxygens (O$_3$).
The bottom panel shows that increasing density yields a reduction in the concentration of Al$_4$ environments, which is compensated by an increase of Al$_6$. In contrast, the Al$_5$ environment displays weaker variations with density; we note that Al$_5$ are characteristic of the amorphous phase of Al$_2$O$_3$ and absent in the crystalline phases \cite{paz_identification_2014}.
Finally, we highlight how the 2.28 g/cm$^3$ model contains a high proportion of O$_2$ bridging oxygens and a majority of Al$_4$ environments, with no Al$_6$. This is consistent with the trend that at lower densities Al becomes primarily tetrahedrally coordinated in Al$_4$ environments \cite{mavric_advanced_2019}. 

From the coexistence of at least five different coordination environments in all the am-Al$_2$O$_3$ models studied, we expect strong structural disorder. To validate these expectations and characterize atomistic disorder, we plot in Fig.~\ref{fig:al2o3_rdf} the total radial distribution function (RDF) for all AIMD models. We highlight how the the total RDF converges to one over a distance shorter than 5 \AA, indicating the presence of strong short-range disorder (\textit{i.e.} within the second coordination shell \cite{elliot_mro,young_probing_2020}).
 The decomposition of the total RDF into partial RDFs will be discussed later in Fig~\ref{fig:al2o3_rdf_mace}, where we use structures containing up to 7680 atoms to show that the oscillations in the partial RDF become very weak (negligible) at distances larger than $\sim~10$ \AA.
We note, in passing, that this decay distance is shorter than the linear size of all the 120-atoms structures studied here, and in agreement with that found in previous work \cite{lizarraga_2011_lowdensity}. 
This suggests that atomistic models containing hundreds of atoms are sufficiently large to capture the structural properties of strongly disordered solids \cite{simoncelli_thermal_2022,Fiorentino_2023}. We validate this expectation in Sec.~\ref{sec:size_effects} using the MACE MLP and atomistic models containing up to 7680 atoms.

\subsection{Vibrational properties} 
\label{sub:vibrational_properties}

The strong variations in the coordination topology observed in the different structures of am-Al$_2$O$_3$ (Fig.~\ref{fig:coord}) are intuitively expected to affect the atomic vibrational excitations.
Thus, in this section we investigate quantitatively how the vibrational properties of am-Al$_2$O$_3$ change with the coordination topology. We used density-functional perturbation theory (DFPT) \cite{baroni_phonons_2001} to compute the energy $\hbar\omega_{\bm{q}s}$ of each vibrational mode $\bm{q}s$, and the corresponding displacement pattern $\mathcal{E}_{\bm{q}s}^{b\alpha}$, which describes how atom $b$ moves in direction $\alpha$ when the vibration $\bm{q}s$ is excited. We label vibrational modes using both the wavevector $\bm{q}$ and the mode index $s$ for the sake of generality, keeping in mind that $\bm{q}$ is necessary only in crystals and in finite-size models of disordered solids, while in ``ideal glasses'' (namely, an astronomically large set of atoms whose arrangement lacks long-range order) one would obtain $\bm{q}=\bm{0}$, thus $s$ would be sufficient to label the vibrations \cite{simoncelli_thermal_2022}.
Then, we use these quantities to investigate if there is a relationship between atomic displacements and coordination topology, computing the root mean square displacement of every atom $b$ \cite{wallace1998thermodynamics}, 
\begin{equation}
\label{eq:RMSD}
    \text{RMSD}(b,T){=}\sqrt{\frac{\hbar}{N_c m_b} \sum_{\bm{q}s} \frac{1}{\omega_{\bm{q}s}}\Big[\tfrac{1}{2} + \tenscomp{N}_{\bm{q}s} \Big] \sum_{\alpha} |\mathcal{E}_{\bm{q}s}^{b\alpha}|^2},
    \raisetag{5pt}
\end{equation}
where  $\tenscomp{N}_{\bm{q}s} {=} [{\exp{(\hbar \omega_{\bm{q}s} / k_B T)} {-} 1}]^{-1}$ is the Bose-Einstein distribution at temperature $T$, $m_b$ is the mass of atom $b$, and $N_c$
is a normalization factor\footnote{N is equal to the number of $\bm{q}$ points used to sample the Brillouin Zone; in the case of an ideal glass (astronomically large disordered simulation cell) the Brillouin Zone reduces to the point $\bm{q}=\bm{0}$ only and $N=1$.}.
\begin{figure}[b!]
  \centering
\includegraphics[width=\WidthFigure]{\folder/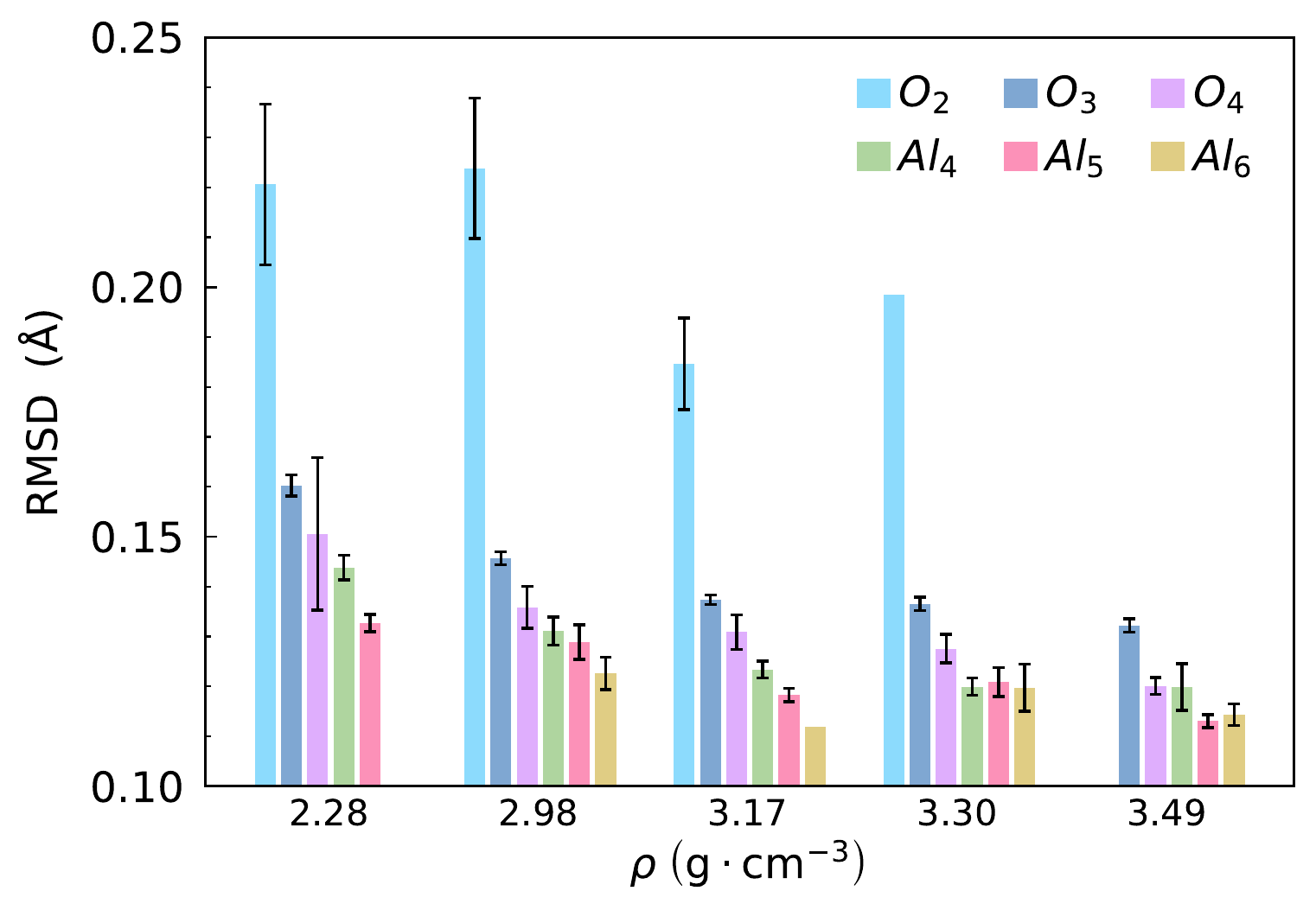}\\[-3mm]
  \caption{\textbf{Average atomic root mean square displacement in am-Al$_2$O$_3$ at various densities,} calculated at  $T=300K$ using Eq.~(\ref{eq:RMSD}).  The average RMSD of oxygen decreases as coordination increases; in contrast, aluminum displays a much weaker anticorrelation between RMSD and coordination number in models with $\rho \leq 3.17$ g/cm$^3$, such an anticorrelation disappears in models with $\rho \geq 3.30$ g/cm$^3$. Error bars represent the standard deviation of the average RMSD, and are not reported in the cases in which only a single coordination environment was detected.}
  \label{fig:MSD_atoms}
\end{figure}

To see if a relationship between RMSD and coordination topology exists, we compute the average ${\rm RMSD}(b,T)$ over atoms with equal coordination. Fig.~\ref{fig:MSD_atoms} shows that the average RMSD of O$_x$ atoms decreases as coordination increases in all the atomistic models analyzed. The average RMSD of Al$_y$ is anticorrelated with coordination in models with $\rho{\leq} 3.17$ g/cm$^3$, and displays deviations from this anticorrelated trend in models with $\rho = 3.30$ g/cm$^3$ (where all Al atoms have a very similar average RMSD, regardless of their coordination) and with
$\rho = 3.49$ g/cm$^3$ (where Al$_6$ has average RMSD slightly larger than Al$_5$).
Interestingly, as density increases the average RMSD of O$_4$ approaches the RMSD of Al$_4$---considering the significant differences between the mass of oxygen and aluminum ($m_{Al}/m_O\approx 1.7$), we note that this behavior departs from the trend  RMSD $\propto \sqrt{{m_b}^{-1}}$  ubiquitously observed in solids with simple coordination topology. This shows that the presence of complex coordination topologies can have non-trivial effects on the vibrational properties.

Having investigated how the constraints imposed by the coordination topology affect the average RMSDs, we now study how coordination influences the vibrational frequencies. 
To this aim, we compute the vibrational density of states (VDOS), $g(\omega){=}({\mathcal{V}N_{\rm c}})^{-1}\sum_{\bm{q},s}\delta(\omega{-}\omega_{\bm{q}s})$ (here, $\mathcal{V}$ is the volume of the cell used to simulate the disordered system), we use the eigenvectors to decompose the VDOS into partial (single-atom) contributions (PDOS), and finally we integrate the single-atom PDOS using an indicator function that allows us to resolve different coordination environments: 
\begin{equation}
    \label{eq:pVDOS}
    g_{t_x}(\omega) = \frac{1}{\mathcal{V} N_c} \sum_{{\bm{q}s}} \delta(\omega - \omega_{\bm{q}s}) \sum_{b, \alpha} |\mathcal{E}^{b\alpha}_{{\bm{q}s}}|^2 \delta_{b , t_x},
\end{equation}
where $\delta_{b , t_x}$ is indicator function equal to one if the atom $b$ is of type $t$ and has coordination $x$ (e.g., $t_x=O_3$ for threefold-coordinated oxygen), and zero otherwise.
Thus, $g_{t_x}(\omega)$ allows us to resolve how the coordination topology affects the VDOS.

\begin{figure}[t]
  \centering
\includegraphics[width=\WidthFigure]{\folder/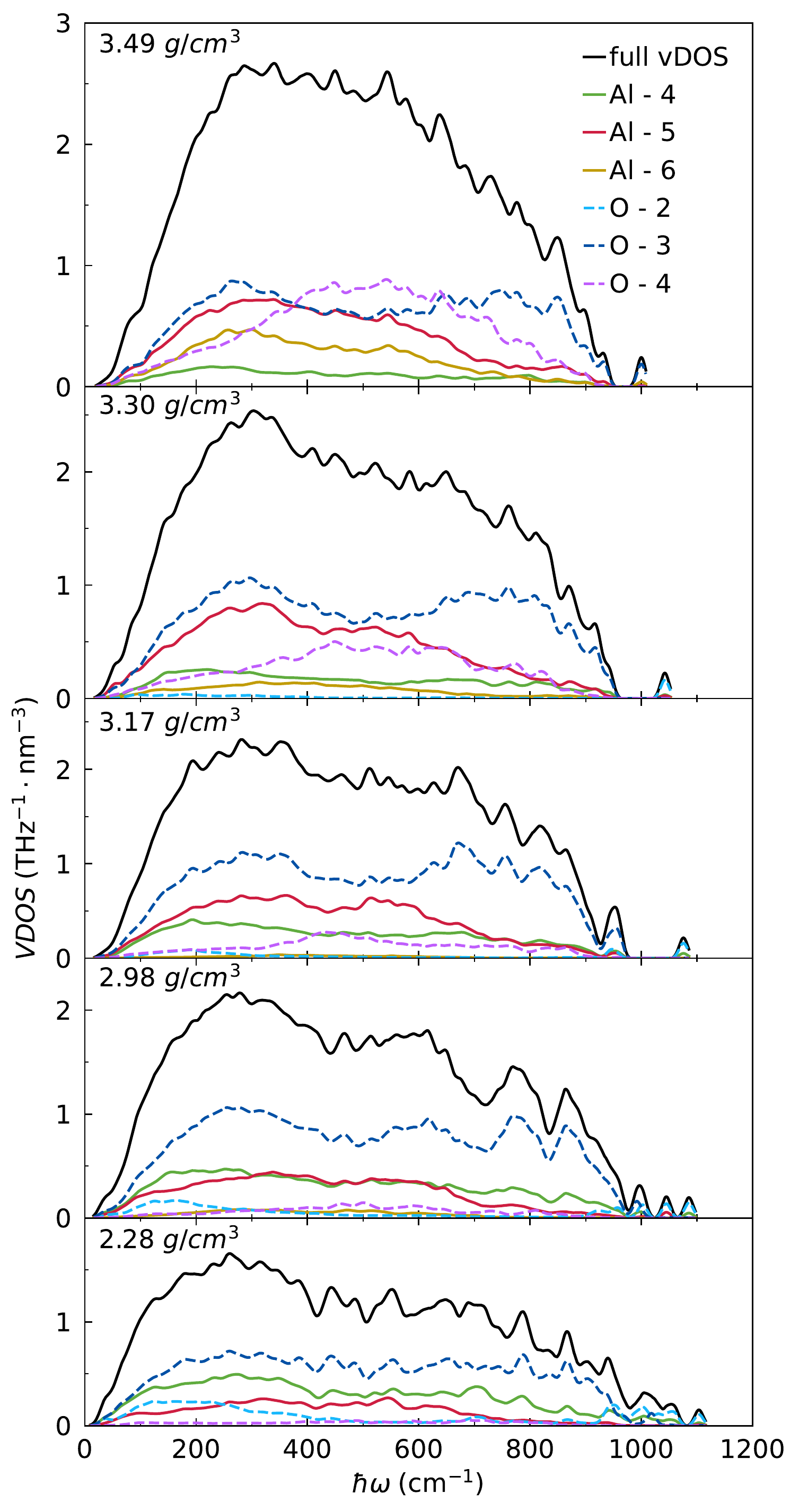}\\[-3mm]
  \caption{
  \textbf{Vibrational density of states of am-Al$_2$O$_3$ at various densities.} The total VDOS is solid black. The colored solid lines distinguish coordination environments for Al atoms: green is Al$_4$, red is Al$_5$, and yellow is Al$_6$. Dashed colored lines are used for coordination environments of O atoms: cyan for O$_2$, blue for O$_3$, and purple for O$_4$.
  }
  \label{fig:VDOS}
\end{figure}
Results for $g_{t_x}(\omega)$ are shown in Fig.~\ref{fig:VDOS}. 
We highlight how the PDOS for different coordination environments have similar shapes at different densities. In particular, different coordination environments for oxygen have fingerprints in the coordination-resolved PDOS: O$_3$ is characterized by a bimodal PDOS with a local minimum around 500 cm$^{-1}$, while O$_2$ and O$_4$ have a unimodal PDOS with peaks at about 150 and 500  cm$^{-1}$, respectively. In contrast, the coordination-resolved PDOS for Al$_4$, Al$_5$, and Al$_6$ are similar in shape. We note that this is in sharp contrast to the crystalline $\theta$-Al$_2$O$_3$ phase, which contains only Al$_4$ tetrahedra and Al$_6$ octahedra and displays a clear distinction between high-frequency breathing modes associated with Al$_4$ environments, and low-frequency bending modes associated with Al$_6$ environments \cite{lodziana2003dynamical}.

Varying the density of am-Al$_2$O$_3$ causes a variation in the relative magnitude of the Al$_x$ PDOS: at 2.28 g/cm$^3$, the Al$_4$ PDOS is always larger in magnitude than Al$_5$; with increasing density, the relative magnitude of the PDOS of Al$_4$ and Al$_5$ progressively reverses, with the highest-density 3.49 g/cm$^3$ model featuring a PDOS for Al$_5$ always larger than that for Al$_4$.

We note that the O$_3$ and Al$_5$ coordination environments have a significant PDOS in all the models analyzed, regardless of the models' density. 
In contrast, the PDOS of O$_2$, O$_4$, Al$_4$, Al$_6$, display much stronger variations with density---increasing density yields a progressive suppression of the O$_2$ and Al$_4$ environments, compensated by the progressive appearance of O$_4$ and Al$_6$ environments.

\section{Thermal properties} 
\label{sec:thermal_properties}

\subsection{Thermal conductivity of glasses} 
\label{ssec:Wigner_formulation}

In this section we summarize the salient features of the Wigner formulation of thermal transport \cite{simoncelli2019unified}, which describes heat conduction accounting for the interplay between structural disorder, anharmonicity, and quantum Bose-Einstein statistics of atomic vibrations. This allows us to describe the thermal conductivity of solids ranging from crystals \cite{PhysRevX.10.011019,simoncelli2021Wigner,Lucente} to glasses \cite{simoncelli_thermal_2022,liu_unraveling_nodate}.

When applied to atomistic models of amorphous systems, the Wigner formulation of thermal transport yields the following `rWTE' conductivity expression\cite{simoncelli_thermal_2022}:
\begin{equation}
\begin{split}
\kappa{=}\frac{1}{\mathcal{V}{N_{\rm c}} }{\sum_{\bm{q},s,s'}}&\!
\frac{\omega_{\bm{q}s}{+}\omega_{\bm{q}s'}}{4}\!\left(\frac{C_{\bm{q}s}}{\omega_{\bm{q}s}}{+}\frac{C_{\bm{q}s'}}{\omega_{\bm{q}s'}}\right)\!
\frac{\rVert\tens{v}(\bm{q})_{s,s'}\lVert^2}{3}\\
\times&\pi\mathcal{F}_{[\eta;\Gamma_{\bm{q}s}{+}\Gamma_{\bm{q}s'}]}(\omega_{\bm{q}s}-\omega_{\bm{q}s'})\;,
\label{eq:thermal_conductivity_combined}
\end{split}
\raisetag{5mm}
\end{equation}
where $\hbar\omega_{\bm{q}s}$ denotes the harmonic energy of the vibration $\bm{q}s$, and $\hbar\Gamma_{\bm{q}s}$ its linewidth (broadening of the harmonic energy level due to anharmonicity~\cite{paulatto2013anharmonic,fugallo2013ab,phono3py,alamode,cepellotti_phoebe_2022,carrete_almabte_2017,kaldo} and isotopic-mass disorder~\cite{tamura_isotope_1983}); 
$\rVert\tens{v}(\bm{q})_{s,s'}\lVert^2{=}\sum_{\alpha=1}^3\tenscomp{v}^{\alpha}(\bm{q})_{s,s'}\tenscomp{v}^{\alpha}(\bm{q})_{s',s}$
is the square modulus of the velocity operator \cite{simoncelli2021Wigner} between eigenstates $s$ and $s'$ at fixed $\bm{q}$ ($\alpha$ denotes a Cartesian direction, and since amorphous solids are in general isotropic, the scalar conductivity~(\ref{eq:thermal_conductivity_combined}) is computed as the average trace of the tensor $\kappa^{\alpha\beta}$ \cite{simoncelli_thermal_2022});
\begin{equation}
C_{\bm{q}s}{=}C[\omega_{\bm{q}s},T]{=}\frac{\hbar^2\omega^2_{\bm{q}s} }{k_{\rm B} T^2} {\tenscomp{N}}_{\bm{q}s}\big({\tenscomp{N}}_{\bm{q}s}{+}1\big)  
\label{eq:quantum_specific_heat_A}
\end{equation}
is the quantum specific heat of the mode $\bm{q}s$; 
$\mathcal{V}$, $N_{\rm c}$, and $\tenscomp{N}_{\bm{q}s}$ are, in order, the simulation's cell volume, normalization factor 
and Bose-Einstein distribution already discussed in Sec.~\ref{sub:vibrational_properties}.
Finally, $\mathcal{F}_{[\eta;\Gamma_{\bm{q}s}{+}\Gamma_{\bm{q}s'}]}(\omega_{\bm{q}s}-\omega_{\bm{q}s'})$ is a Voigt distribution, \textit{i.e.} a two-parameter distribution that reduces to a Lorentzian with full width at half maximum (FWHM) $\Gamma_{\bm{q}s}{+}\Gamma_{\bm{q}s'}$ when $\Gamma_{\bm{q}s}{+}\Gamma_{\bm{q}s'}\gg \eta$, and to a Gaussian representation of the Dirac delta with variance $\eta^2\pi/2$ in the opposite limit ($\Gamma_{\bm{q}s}{+}\Gamma_{\bm{q}s'}\ll \eta$). 

As discussed in detail in Ref.~\cite{simoncelli_thermal_2022}, Eq.~(\ref{eq:thermal_conductivity_combined}) describes thermal transport as originating from couplings between vibrations that have an energy difference smaller than the broadening of the Voigt profile. Such a broadening is determined by $\eta$ in the low-temperature (harmonic) limit where anharmonicity phases out ($\Gamma_{\bm{q}s}{+}\Gamma_{\bm{q}s'}\to 0\;\forall\;\bm{q},s$). Setting $\eta$ to a value slightly larger than the average energy-level spacing $\hbar\Delta\omega_{\rm avg}=\frac{\hbar\omega_{\rm max}}{3\cdot N_{\rm at}}$ ($\hbar\omega_{\rm max}$ is the maximum vibrational energy, $3N_{\rm at}$ is the number of vibrational energy levels, \textit{i.e.} three times the number of atoms in the system's reference cell) accounts for the physical property that heat transfer via a wave-like tunneling between neighboring (quasi-degenerate) vibrational eigenstates can occur~\footnote{A necessary condition for this to happen is to have vibrations that are not localized in the Anderson sense \cite{Anderson_localization}, \textit{i.e.} to have non-zero velocity operator elements in Eq.~(\ref{eq:thermal_conductivity_combined}).} even in the limit of vanishing anharmonicity, implying that in such a limit Eq.~(\ref{eq:thermal_conductivity_combined}) reduces to the harmonic Allen-Feldman (AF) thermal conductivity for glasses \cite{allen1989thermal,allen1993thermal}.
In contrast, at temperatures where the anharmonic linewidths are much larger than the computational broadening $\eta$, the Voigt profile becomes a Lorentzian with FWHM $\Gamma_{\bm{q}s}{+}\Gamma_{\bm{q}s'}$, effectively reducing to the anharmonic Wigner conductivity expression~\cite{simoncelli2021Wigner}.
In practice, when applied to finite-size atomistic models of amorphous solids at temperatures for which $\hbar\Delta\omega_{\rm avg}> \Gamma_{\bm{q}s}$, the rWTE conductivity expression~(\ref{eq:thermal_conductivity_combined})  ensures that the low-temperature harmonic Allen-Feldman limit is correctly described, and the effects of anharmonicity are considered only when they are not spuriously altered by finite-size effects \cite{simoncelli2021Wigner}.  

In actual calculations, an amorphous solid is approximately described as a crystal having a primitive cell containing a large but finite number of atoms $N_{\rm at}$. Thus, the Brillouin Zone (BZ) corresponding to the (large) finite-size model does not reduce to the aforementioned ``ideal glass'' limit ($\bm{q}{=}\bm{0}$ only), but has a (small) finite volume.
Recent work \cite{simoncelli_thermal_2022} has shown that when the lengthscale of structural disorder is shorter than the size of the simulation cell, periodic boundary conditions and Fourier interpolation over the small BZ of the glass can be exploited to improve the sampling of the vibrational properties, and extrapolate the bulk limit of the thermal conductivity. 
In practice, using the Fourier interpolation in a disordered model corresponds to averaging over different possible boundary conditions. For example, the vibrational properties computed at $\bm{q}{=}\bm{0}$ only correspond to considering periodic boundary conditions at the boundaries of the simulation box; calculations at $\bm{q}{=}(1, 0, 0)\pi/L$ correspond to considering antiperiodic boundary conditions along direction $x$ and periodic along the other directions (\citet{feldman_thermal_1993}). 
Of course, increasing the size of the atomistic model, one expects the choice of the boundary conditions to become less and less relevant, and in practice this can be verified by comparing a calculation at $\bm{q}{=}\bm{0}$ only with one performed relying on Fourier interpolation---when differences are negligible, one can conclude that the boundary conditions are irrelevant and the system is large enough to allow a brute-force simulation of the bulk limit.
In practice it is not always possible to simulate this, due to exceedingly large computational cost. In the next section we show that in finite-size models of strongly disordered systems such as am-Al$_2$O$_3$, the  bulk limit of the conductivity can be computed using the rWTE~(\ref{eq:thermal_conductivity_combined}), which ensures that: (i) heat transfer mediated by tunneling between neighboring eigenstates can always occur, preserving a physical property that would otherwise emerge only in the thermodynamic limit; (ii) the sampling of the vibrational properties is improved by averaging over different boundary conditions.
These statements are validated in Sec.~\ref{sec:size_effects} using the MACE MLP for a brute-force calculation of the bulk limit in models containing up to 7680 atoms.

Finally, it is worth mentioning that the Wigner thermal conductivity expression~(\ref{eq:thermal_conductivity_combined}) can be derived also from a many-body Green-Kubo approach~\cite{caldarelli_many-body_2022,PhysRevB.107.054311}, and such an expression has been recently employed~\cite{isaeva2019modeling,lundgren_mode_2021}, in combination with interatomic potentials, to describe the thermal properties of several glasses \footnote{More precisely , Refs.~\cite{isaeva2019modeling,lundgren_mode_2021} evaluated the bulk limit of the conductivity relying on empirical interatomic potentials and atomistic models containing thousands of atoms, \textit{i.e.}  having a size large enough to achieve computational convergence by evaluating Eq.~(\ref{eq:thermal_conductivity_combined}) at $\bm{q}=\bm{0}$ only and without relying on the Voigt regularization. See Ref.~\cite{simoncelli_thermal_2022} for details on the conditions under which evaluating Eq.~(\ref{eq:thermal_conductivity_combined}) yields equivalent results with or without relying on the Fourier interpolation and Voigt regularization.}.

\subsection{Numerical results} 
\label{sub:numerical_results}
\subsubsection{Thermal conductivity from first principles} 
\label{ssub:thermal_conductivity}

In this section we evaluate the rWTE conductivity from first principles (see Appendix~\ref{sec:computational_details} for details) for each am-Al$_2$O$_3$ structure, in the temperature range from 50\,K to 700\,K. We focus on this temperature range because it is the most relevant for technological applications related to electronics \cite{jin2019li4ti5o12,qu2019air,randau2021additive}, and can be studied with the 120-atom models at our disposal \cite{simoncelli_thermal_2022} (this last statement is discussed in detail and validated in Sec.~\ref{sec:size_effects}, where we use a MLP to study models containing up to 7680 atoms).

\begin{figure}[b]
  \centering
\includegraphics[width=\WidthFigure]{\folder/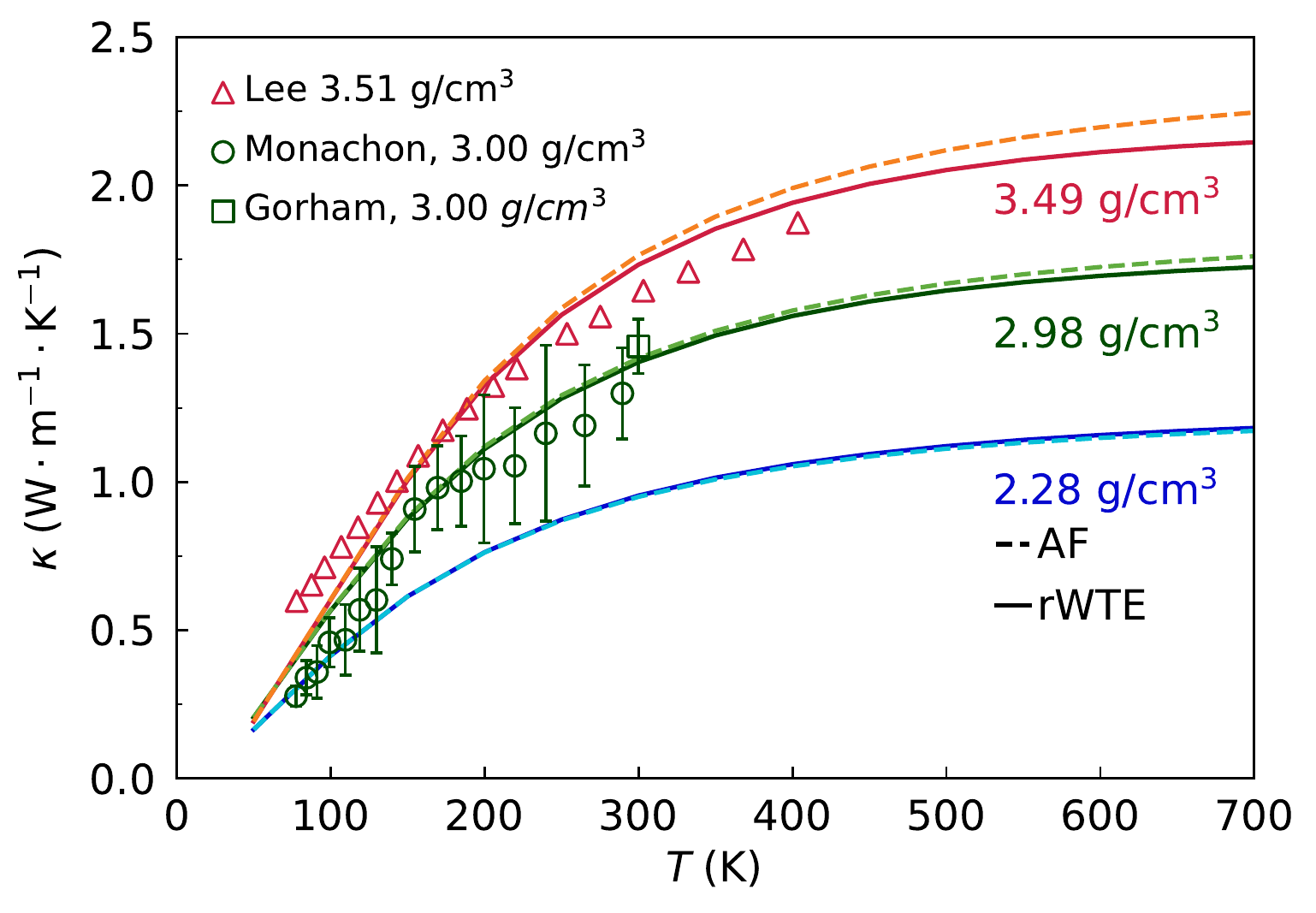}
  \caption{\textbf{Wigner vs Allen-Feldman thermal conductivity of am-Al$_2$O$_3$}.
  Predictions from anharmonic rWTE are solid, from harmonic AF are dashed; different colors distinguish different densities 2.28 g/cm$^3$ (blue and cyan), 2.98 g/cm$^3$ (dark and light green) and 3.49 g/cm$^3$ (red and orange).  
  The effects of anharmonicity are overall small, and become weaker as density decreases. 
  Scatter points are experiments: red triangles are taken from \citet{Lee1995} (DC sputtering), green circles from \citet{Monachon2015} (ALD on Si substrate), and the green square is from \citet{gorham_density_2014} (ALD on Si substrate). 
 Theory and experiments are in reasonably good agreement over the entire temperature range.}
  \label{fig:Wigner_vs_AllenFeldman_tc}
\end{figure}
We show in Fig.~\ref{fig:Wigner_vs_AllenFeldman_tc} the predictions for the thermal conductivity obtained using the harmonic AF or anharmonic rWTE formulations (see Appendix~\ref{sec:convergence_of_the_allen} for details on the convergence test for the AF broadening parameter $\eta$ and for the calculation of the anharmonic linewidths). We find that the effects of anharmonicity are in general weak---AF and rWTE differ at most by 10$\%$ in the highest-density (3.49\,g/cm$^3$) model. These differences become smaller as the density decreases, and are practically invisible in the lowest-density (2.28\,g/cm$^3$) model.
The good agreement between AF and rWTE shows that in am-Al$_2$O$_3$ the vibrations' damping due to disorder is strong enough to dominate over anharmonicity. In addition, we highlight how both AF and rWTE predict an increasing-up-to-saturation trend for the temperature-conductivity curve; such a saturating trend is inherited from that of the specific heat (more on this later in Sec.~\ref{ssub:diffusivity}).
Our calculations shed light on the thermal properties of am-Al$_2$O$_3$ below room temperature, where the quantum Bose-Einstein statistics of vibrations has a major effect on thermal transport. This is a step forward compared to previous studies based on molecular dynamics (MD) \cite{li_effects_2020}, which were governed by classical equipartition and thus had to be limited to the high-temperature regime \cite{PhysRevMaterials.3.085401} (\textit{i.e.} at temperature large enough to have a quantum specific heat effectively very close to the constant classical limit).

In Fig. \ref{fig:Wigner_vs_AllenFeldman_tc} we compare our calculations with the experimental measurements from Refs.~\cite{Monachon2015,gorham_density_2014} (ALD samples grown on Si substrate and having density 3.0\,g/cm$^3$) and Ref.~\cite{Lee1995} (DC-sputtered samples with density 3.51\,g/cm$^3$).
Thermal conductivity experiments have uncertainties that depend on many factors, including, e.g., the sample’s size, shape, and measurement method \cite{fournier_measurement_2020,koh_frequency_2007,zhou_2020_thermal}; as shown by the error bars from Refs.~\cite{Monachon2015,gorham_density_2014}, these uncertainties are typically around $\sim 10-20\%$. 
We see that our predictions at different densities are overall compatible (within the error bars or within 10$\%$) with the corresponding experiments.

\begin{figure}[t]
  \centering
  \includegraphics[width=\WidthFigure]{\folder/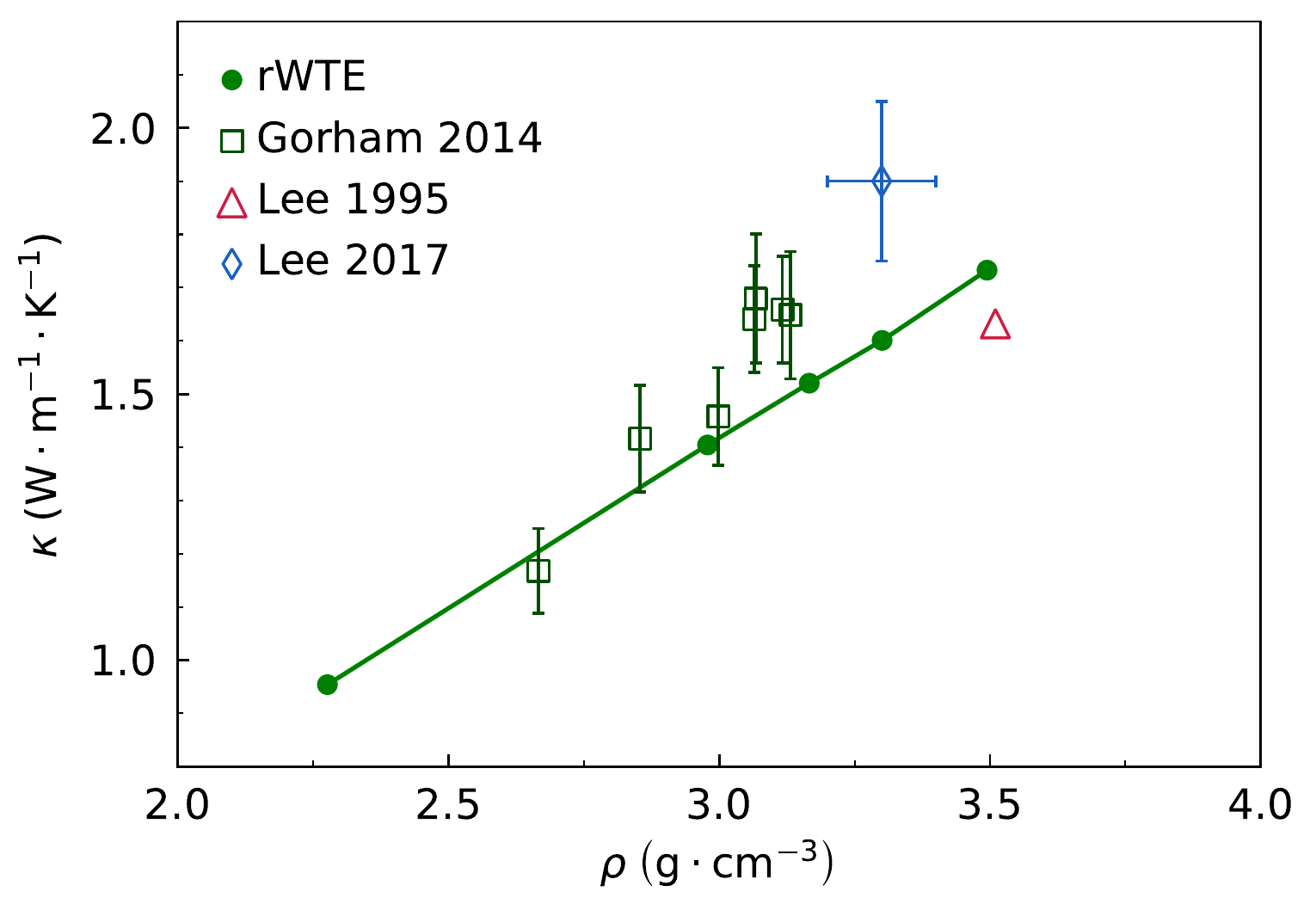}
  \caption{
\textbf{Thermal conductivity as a function of density.} 
  The solid green line is the theoretical rWTE conductivity computed at 300 K. 
  Empty scatter points are experiments performed at 300 K by \citet{gorham_density_2014} in 2014 (green squares, ALD samples grown on Si), by \citet{Lee2017} in 2017 (blue diamond, ALD samples grown on Si), and by \citet{Lee1995} in 1995 (red triangle, DC sputtering).}
  \label{fig:k_vs_density}
\end{figure}

We now turn our attention to the dependence of the conductivity on density. 
In Fig.~\ref{fig:k_vs_density} we plot the room-temperature rWTE conductivity as a function of density (we note from Fig.~\ref{fig:Wigner_vs_AllenFeldman_tc} that at 300 K the rWTE and AF conductivities are practically indistinguishable across the entire density range analyzed), finding an approximately linear increase of the conductivity with density ($\kappa(\rho)_{300K}\approx a\cdot\rho +b$, where $a {=} 0.637 {\pm} 0.004 \tfrac{\rm W\cdot cm^3}{\rm m\cdot K \cdot g}$, $b {=} {-}0.495 {\pm}0.014 \tfrac{\rm W}{\rm m\cdot K }$). 
We compare our predictions with the experiments by \citet{gorham_density_2014}, who characterized the room-temperature thermal conductivity of ALD am-Al$_2$O$_3$ grown on Si substrate~\footnote{We note, in passing, that Refs. \cite{Monachon2015,Lee2017,gorham_density_2014} did not observe a significant dependence of the thermal conductivity from the substrate on which the am-Al$_2$O$_3$ sample was grown.} at densities ranging from 2.66 g/cm$^3$ to 3.12 g/cm$^3$. 
To give an idea about the conductivity variability found when comparing independent experiments, we also report the room-temperature conductivities extracted from the dataset of \citet{Lee1995} (1995, DC-sputtered) and \citet{Lee2017} (ALD on Si, 2017) already presented in Fig.~\ref{fig:Wigner_vs_AllenFeldman_tc}.
We note that there is some variance between the conductivity observed in independent experiments, still they show a conductivity that overall increases with density, and such a trend is reproduced in our calculations.

\subsubsection{Thermal diffusivity and effects of anharmonicity} 
\label{ssub:diffusivity}

To gain microscopic, fundamental insights on why anharmonicity has negligible effects on the thermal conductivity of am-Al$_2$O$_3$, it is useful to resolve the amount of heat carried by each atomic vibration and its diffusion rate. 
This information can be obtained by recasting the rWTE conductivity expression~(\ref{eq:thermal_conductivity_combined}) as \cite{simoncelli_thermal_2022}
\begin{equation}
  \kappa(T)=\int_{0}^{\omega_{\rm max}} g(\omega) C(\omega,T) D(\omega,T) d\omega\;,
  \label{eq:kappa_diff}
\end{equation}
where $\omega_{\rm max}$ is the maximum vibrational frequency of the system, $g(\omega)$ is the VDOS discussed in Sec.~\ref{sub:vibrational_properties}), $C(\omega,T)$ is the specific heat for a vibration with frequency $\omega$ at temperature $T$ (see Eq.~(\ref{eq:quantum_specific_heat_A})), and $D(\omega,T)$ is the rWTE diffusivity \cite{simoncelli_thermal_2022}, 
\begin{equation}
  D(\omega,T){=}[g(\omega){\mathcal{V}N_{\rm c}}]^{-1}\sum_{\bm{q},s} D_{\bm{q}s} \delta(\omega{-}\omega_{\bm{q}s}),\label{eq:diff_omega}
\end{equation}
\begin{align}
D_{\bm{q}s}{=}&{\sum_{s'}}
\frac{\omega_{\bm{q}s}{+}\omega_{\bm{q}s'} }{{2[C_{\bm{q}s}{+}C_{\bm{q}s'}] }}
\!\left[\!\frac{C_{\bm{q}s}}{\omega_{\bm{q}s}}{+}\frac{C_{\bm{q}s'}}{\omega_{\bm{q}s'}}\!\right]\!\!
\frac{\lVert\tens{v}(\bm{q})_{s,s'}\lVert^2\!}{3}\label{eq:diffusivity_q_s}\\
&\hspace{5mm}\times\pi\mathcal{F}_{[\eta;\Gamma_{\bm{q}s}{+}\Gamma_{\bm{q}s'}]}(\omega_{\bm{q}s}-\omega_{\bm{q}s'})\;.\nonumber
\end{align}
The expression for $D_{\bm{q}s}$ is determined by factorizing the specific heat $C_{\bm{q}s}$ in Eq.~(\ref{eq:thermal_conductivity_combined}) and by the requirement that in the coupling between two vibrations ${\bm{q}s}$ and ${\bm{q}s'}$ each contributes to the coupling with a weight equal to the relative specific heat \cite{simoncelli2021Wigner} (e.g. for vibration ${\bm{q}s}$ the weight is $\tfrac{C_{\bm{q}s}}{C_{\bm{q}s}+C_{\bm{q}s'}}$, and correspondingly for vibration ${\bm{q}s'}$ the weight is $\tfrac{C_{\bm{q}s'}}{C_{\bm{q}s}+C_{\bm{q}s'}}$).
In the harmonic AF limit $\eta{\gg}\Gamma_{\bm{q}s}{+}\Gamma_{\bm{q}s'}{\to} 0$, thus the Voigt distribution in Eq.~(\ref{eq:diffusivity_q_s}) reduces to the Gaussian representation of the Dirac $\delta$, accounting only for couplings between quasi-degenerate vibrational eigenstates and effectively reducing to the temperature-independent AF diffusivity~\cite{allen1989thermal,allen1993thermal}. 
We note that in the context of Eq.~(\ref{eq:kappa_diff}) the only difference between AF and rWTE originates from the diffusivity $D(\omega, T)$, implying that differences between AF and rWTE conductivities derive exclusively from differences between the AF and rWTE diffusivities.

\begin{figure}[t]
    \centering
    \includegraphics[width=\WidthFigure]{\folder/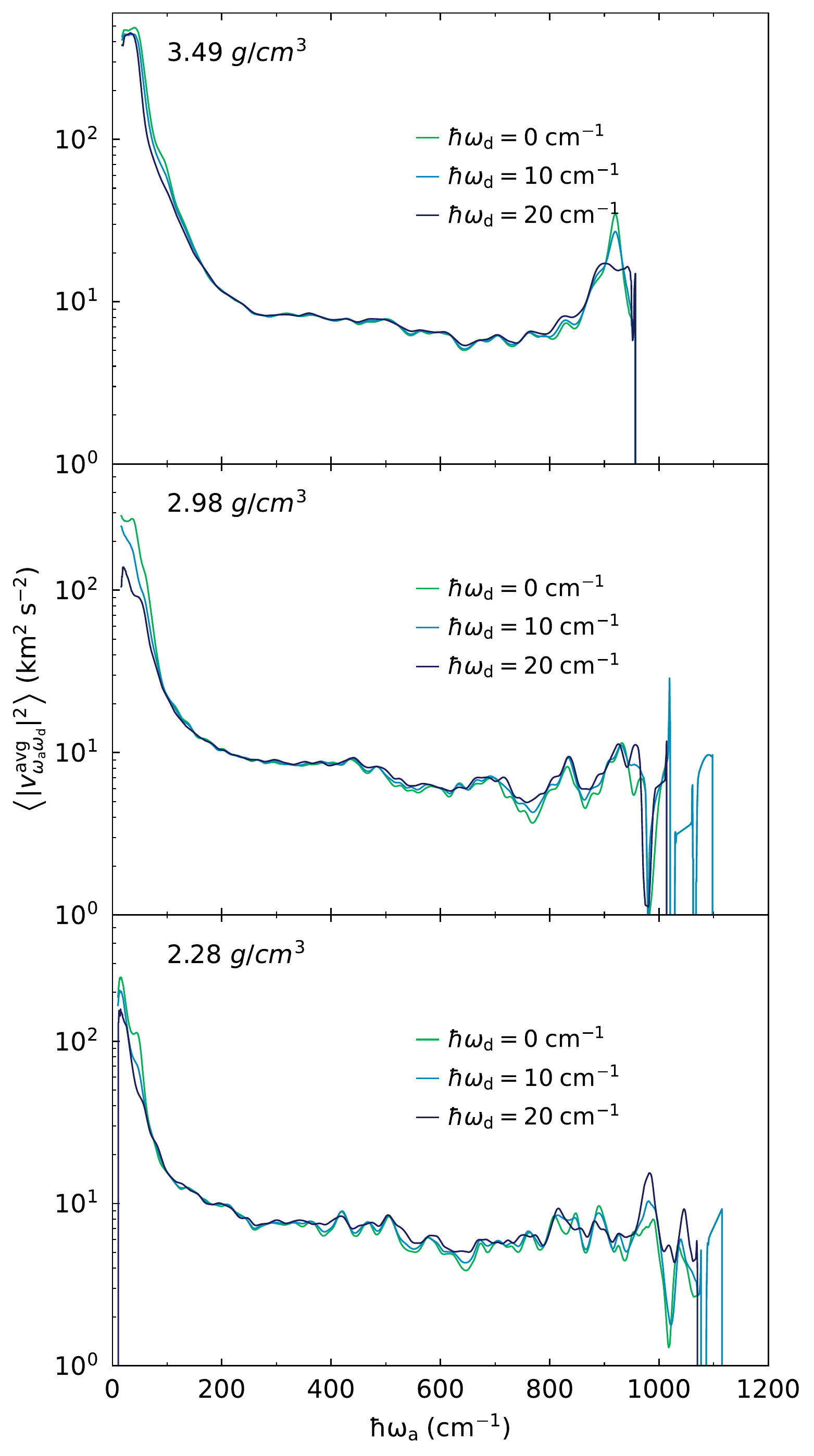}\\[-4mm]
    \caption{
    \textbf{Velocity operator of am-Al$_2$O$_3$ as a function of vibrational energy differences and averages,} for the models having density 2.28, 2.98, and 3.49 g/cm$^3$, computed from Eq.~(\ref{eq:v_operator_omega_a_omega_d}). We see that the average velocity operators for all densities are relatively unchanged at increasing differences, $\omega_d$. These small differences imply that the effects of anharmonicity on the conductivity are negligible (see text).
    }
    \label{fig:vel_op_all_2D}
\end{figure}

Eqs.~(\ref{eq:diff_omega},\ref{eq:diffusivity_q_s}) allow us to obtain additional, microscopic information on why the harmonic AF and anharmonic rWTE conductivities in Fig.~\ref{fig:Wigner_vs_AllenFeldman_tc} display small (negligible) differences in the high-temperature limit ($C_{\bm{q}s}=C(\omega_{\bm{q}s},T){\to} k_B$ and $\Gamma_{\bm{q}s}{\gg} \eta$ $\forall\; \bm{q}s$).
Specifically, the Voigt distribution in Eqs.~(\ref{eq:diff_omega},\ref{eq:diffusivity_q_s}) implies that in the high-temperature limit the rWTE diffusivity is a Lorentzian-weighted average of velocity-operator elements $\lVert\tens{v}(\bm{q})_{s,s'}\lVert^2\!$,  and in practice such average is mainly determined by elements satisfying $|\omega_{\bm{q}s}{-}\omega_{\bm{q}s'}|{<}[\Gamma_{\bm{q}s}{+}\Gamma_{\bm{q}s'}]$; increasing temperature implies an increase of the linewidths, and therefore velocity-operator elements with increasingly larger frequency difference contribute to such average. 
Therefore, the trend of the velocity-operator elements as a function of the energy difference $\hbar|\omega_{\bm{q}s}{-}\omega_{\bm{q}s'}|$ determines the trend of the diffusivity as a function of temperature: elements increasing (decreasing) with frequency difference imply a conductivity increasing (decreasing) with temperature. 
In contrast, in the harmonic AF limit, the diffusivity is always determined by quasi-degenerate velocity-operator elements ($\hbar|\omega_{\bm{q}s}{-}\omega_{\bm{q}s'}|\to 0$) and therefore it does not depend on temperature.
Applying this reasoning to am-Al$_2$O$_3$, where the rWTE conductivity saturates to a temperature-independent value at high temperature, we expect the velocity-operator elements to be roughly independent from the frequency difference. To validate this reasoning, we show in Fig.~\ref{fig:vel_op_all_2D} the velocity operator as a function of the energy average $\hbar\omega_a=\hbar[\omega_{\bm{q}s}{+}\omega_{\bm{q}s'}]/2$ and difference $\hbar\omega_d=\hbar|\omega_{\bm{q}s}{-}\omega_{\bm{q}s'}|$ of the modes coupled \cite{simoncelli_thermal_2022}:
\begin{equation}
\begin{split}
  &\big<|\tenscomp{v}^{\rm avg}_{\omega_{\rm a}\omega_{\rm d}}|^2\big>=[\mathcal{G}(\omega_{\rm a},\omega_{\rm d})]^{-1}\frac{1}{\mathcal{V}N_{\rm c}}{\sum_{\bm{q},s,s'}} \frac{\rVert\tens{v}(\bm{q})_{s,s'}\rVert^2}{3}\\
  &\hspace*{1cm}{\times} 
 \delta\left(\omega_{\rm d}{-}(\omega(\bm{q})_s{-}\omega(\bm{q})_{s'})\right)\delta\Big(\omega_{\rm a}{-}\frac{\omega(\bm{q})_s{+}\omega(\bm{q})_{s'}}{2} \Big)\;;
 \raisetag{15mm}
  \label{eq:v_operator_omega_a_omega_d}
\end{split}
\end{equation}
where $\mathcal{G}(\omega_{\rm a},\omega_{\rm d})$ is a density of states that serves as normalization
\begin{equation}
\begin{split}
  \mathcal{G}(\omega_{\rm a},\omega_{\rm d})
  =\frac{1}{N_{\rm at}}\frac{1}{\mathcal{V}N_{\rm c}}&{\sum_{\bm{q},s,s'}}{\delta}\left(\frac{\omega(\bm{q})_s+\omega(\bm{q})_{s'}}{2}-\omega_{\rm a}\right) \\
 \times&{\delta}\big((\omega(\bm{q})_{s}{-}\omega(\bm{q})_{s'})-\omega_{\rm d}\big)\;.
  \label{eq:2freq_vDOS}
  \raisetag{5mm}
\end{split}
\end{equation}
The plots confirm our expectations, \textit{i.e.}, in the am-Al$_2$O$_3$ models studied the saturating trend of the rWTE conductivity, and the negligible effect of anharmonicity, derive from having microscopic velocity-operator elements that do not vary appreciably with $\omega_{\rm d}$ across the range of densities studied.
Keeping the negligible differences between AF and rWTE diffusivity in mind, in the following we focus on the temperature-\textit{independent} AF limit of the diffusivity $D(\omega)$ to simplify the discussion (\textit{i.e.} Eqs.~(\ref{eq:diff_omega},\ref{eq:diffusivity_q_s}) evaluated with $\Gamma_{\bm{q}s}=0\;\forall \bm{q}s$ and $\eta$ determined from the convergence plateau as discussed in the Appendix \ref{sec:convergence_of_the_allen}).

Eq.~(\ref{eq:kappa_diff}) shows that the contributions to heat transport of atomic vibrational modes with frequency $\omega$ is determined by their density of states $g(\omega)$, amount of heat carried $C(\omega,T)$ and rate of diffusion $D(\omega)$.
In previous sections we showed that: (i) VDOS increases with density as discussed in Fig.~\ref{fig:VDOS}; (ii) conductivity increases (linearly) with density as evidenced in Fig.~\ref{fig:k_vs_density}.
Therefore, it is natural to ask to what extent the conductivity increase observed in Fig.~\ref{fig:k_vs_density} derives from the increase in the VDOS with density, and how density affects the diffusivity of vibrations.  
\begin{figure}[b]
  \centering
\includegraphics[width=\WidthFigure]{\folder/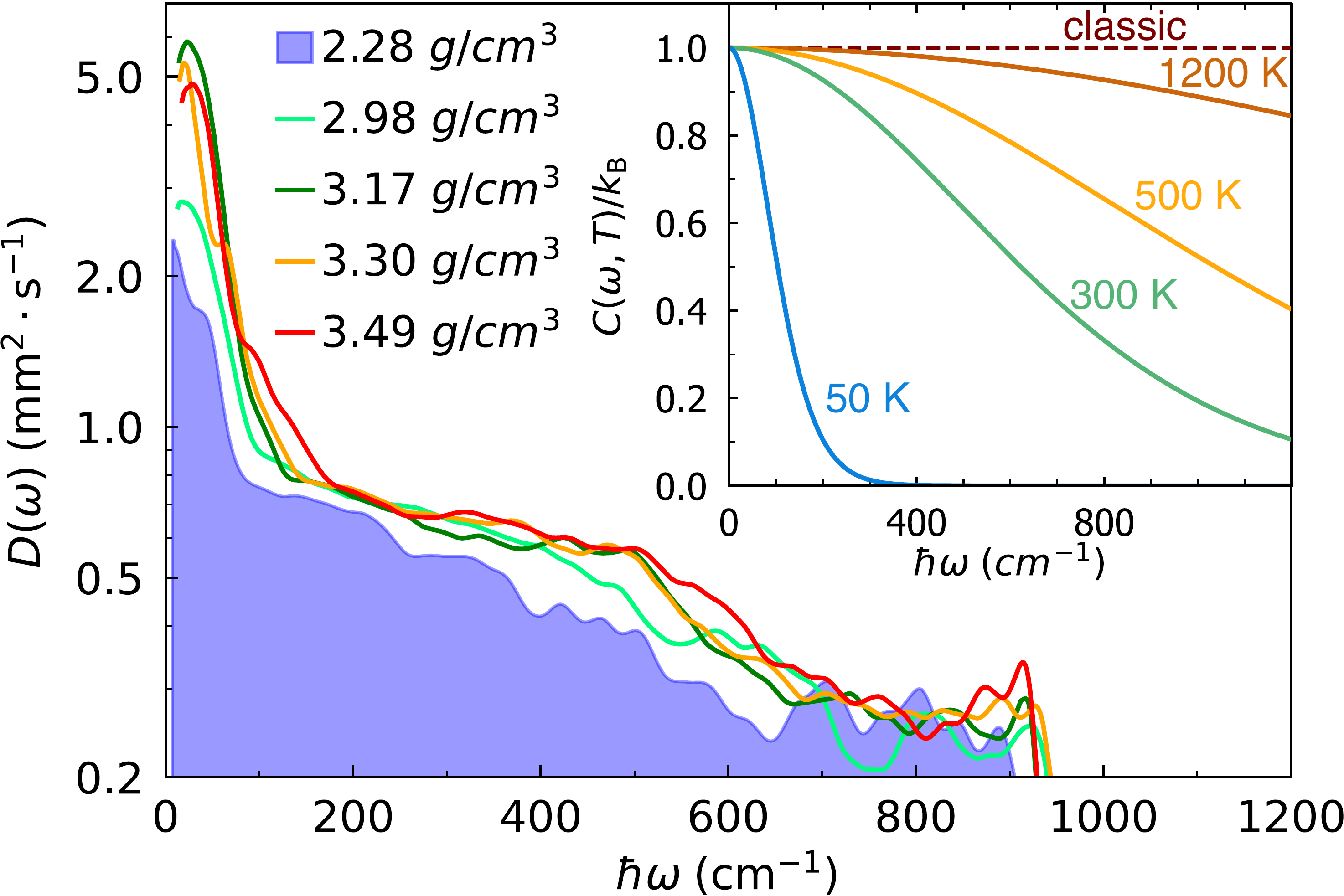}
  \caption{\textbf{AF diffusivity of am-Al$_2$O$_3$ at various densities,} computed using Eq.~\ref{eq:diff_omega} in the AF limit. 
  The diffusivity tends to increase with density, especially at low frequencies (0 - 100 cm$^{-1}$ range) and between 400 and 600 ${\rm cm}^{-1}$. Inset, quantum specific heat as a function of temperature.}
  \label{fig:diffusivity}
\end{figure}
Analyzing the frequency-resolved AF diffusivity $D(\omega)$ reported in Fig.~\ref{fig:diffusivity} for all our models of am-Al$_2$O$_3$ allows us to address these questions. 
We see that an overall increase of diffusivity with density is visible when comparing low-density ($\rho=$2.28 g/cm$^3$), medium-density ($\rho=$2.98 g/cm$^3$) and high-density ($\rho\geq 3.17$ g/cm$^3$) models, especially at low frequencies (from 0 to 100 cm$^{-1}$). 
More precisely, comparing the diffusivity of the highest-density $\rho= 3.49$ g/cm$^3$ model with that of the lowest-density $\rho= 2.28$ g/cm$^3$ model, we find that the highest-density model has a diffusivity that is a factor $\sim$2 larger than that of the lowest-density model in the low-frequency region (0-100 cm$^{-1}$) and a up to a factor 1.5 larger in the region between 400 and 600 cm$^{-1}$. The inset of  Fig.~\ref{fig:diffusivity} shows the quantum specific heat $C(\omega,T)$ at various temperatures, whose frequency dependence at fixed temperature $T$ is indicative of the portion of vibrational spectrum that significantly contributes to heat transport at that temperature.

We highlight how the low-frequency vibrational modes which have density-dependent diffusivity are significantly populated at all the temperatures considered, implying that the increase of the thermal conductivity with density observed in Fig.~\ref{fig:k_vs_density} receives contributions also from increases in the diffusivity.
We also note that the saturation of the specific heat shown in the inset of Fig.~\ref{fig:diffusivity} drives the saturation of the AF thermal conductivity at high temperature (Fig.~\ref{fig:Wigner_vs_AllenFeldman_tc}). Given that the effects of anharmonicity are unimportant for thermal transport in am-Al$_2$O$_3$, the saturation of the rWTE conductivity has to be attributed to the saturation of the specific heat.
Finally, we highlight how the classical limit (dashed line in the inset of Fig.~\ref{fig:diffusivity}) is not yet reached even at temperatures as high as 1200 K; this underscores the importance of correctly accounting for the quantum Bose-Einstein statistics of vibrations to describe thermal transport in am-Al$_2$O$_3$ at technologically relevant temperatures.

\begin{figure}[h]
  \centering
\includegraphics[width=\WidthFigure]{\folder/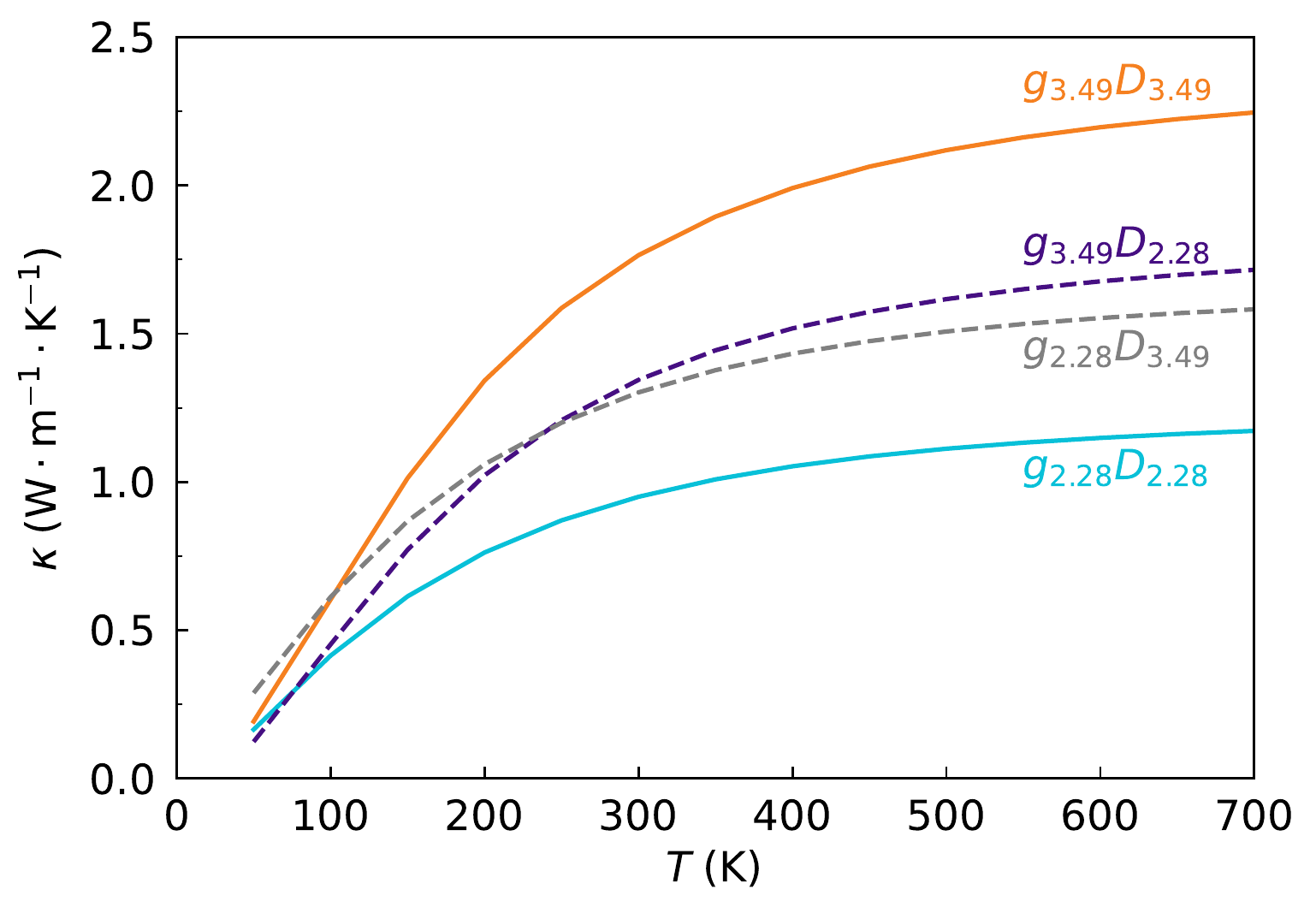}
  \caption{\textbf{Microscopic mechanisms underlying conductivity increase with density.}
  Solid lines are exact AF conductivity calculation for the highest-density 3.49 g/cm$^3$ model (orange, $g_{3.49} D_{3.49}$) and lowest-density 2.28 g/cm$^3$ model (cyan, $g_{2.28} D_{2.28}$).
  The dashed lines show results of artificial conductivity calculations, performed using Eq.~(\ref{eq:kappa_diff}) 
  with 
  the VDOS of the highest-density model and the diffusivity of the lowest-density model (purple, $g_{3.49} D_{2.28}$),  or the VDOS of the  lowest-density model and the diffusivity of the highest-density model (grey, $g_{2.28} D_{3.49}$).
  The artificial calculations yield conductivities lying approximately halfway between the exact limits, indicating that 
  the increase of conductivity with density is determined, in similar proportion, by both an increases in vDOS and by an increase in diffusivity.}
  \label{fig:artificial_conductivity}
\end{figure}
In summary, we have found that both VDOS (Fig.~\ref{fig:VDOS}) and diffusivity (Fig.~\ref{fig:diffusivity}) increase with density.
In order to estimate how much the increase of conductivity with density shown in Figs.~\ref{fig:Wigner_vs_AllenFeldman_tc},\ref{fig:k_vs_density} depends on increase of the VDOS and how much on the increase in diffusivity, we computed the conductivity  artificially combining the VDOS and diffusivity of the highest and lowest-density models in the following ways: (i) using the VDOS of the 3.49 g/cm$^3$ model and the AF diffusivity of the 2.28 g/cm$^3$ model; (ii) using the VDOS of the 2.28 g/cm$^3$ model and the AF diffusivity of the 3.49 g/cm$^3$ model. The comparison between these artificial conductivities and the exact ones (taken from Fig.~\ref{fig:Wigner_vs_AllenFeldman_tc}) are reported in Fig.~\ref{fig:artificial_conductivity}, and show that 
these two artificial calculations yield
a conductivity that lies approximately halfway between the actual conductivities of the lowest-density and highest-density models over a wide temperature range, demonstrating that variations of the thermal conductivity with density are determined by both variations in the VDOS and in the diffusivity.

\section{Size effects}
\label{sec:size_effects}

The AF and rWTE conductivities presented in the previous section were evaluated following the protocol presented in Ref.~\cite{simoncelli_thermal_2022}, which showed that the thermal conductivity of strongly disordered solids such as vitreous silica can be accurately reproduced using models containing ${\sim}100$ atoms, a result supported by other recent studies \cite{ndour_practical_2023,Fiorentino_2023}. 
In this section we study how the size of the atomistic model affects the theoretical predictions for the conductivity of am-Al$_2$O$_3$.
To overcome the limitations posed by the computational cost of first-principles calculations, we generated a MACE MLP \cite{batatia2022mace,kovacs2023evaluation} for am-Al$_2$O$_3$ using the first-principles dataset released by Li \textit{et al.} \cite{li_effects_2020}; computational details are reported in Appendix~\ref{sub:fit_of_the_mace_machine_learning_potential}. Then, we used this potential to produce am-Al$_2$O$_3$ models containing 120, 3240, and 7680 atoms before finally computing the vibrational properties of these models, as well as the AF and rWTE conductivities.

\subsection{Model generation via melt-quench}
\begin{figure}[b]
  \centering
  \hspace*{2mm} \includegraphics[width=0.85\WidthFigure]{\folder/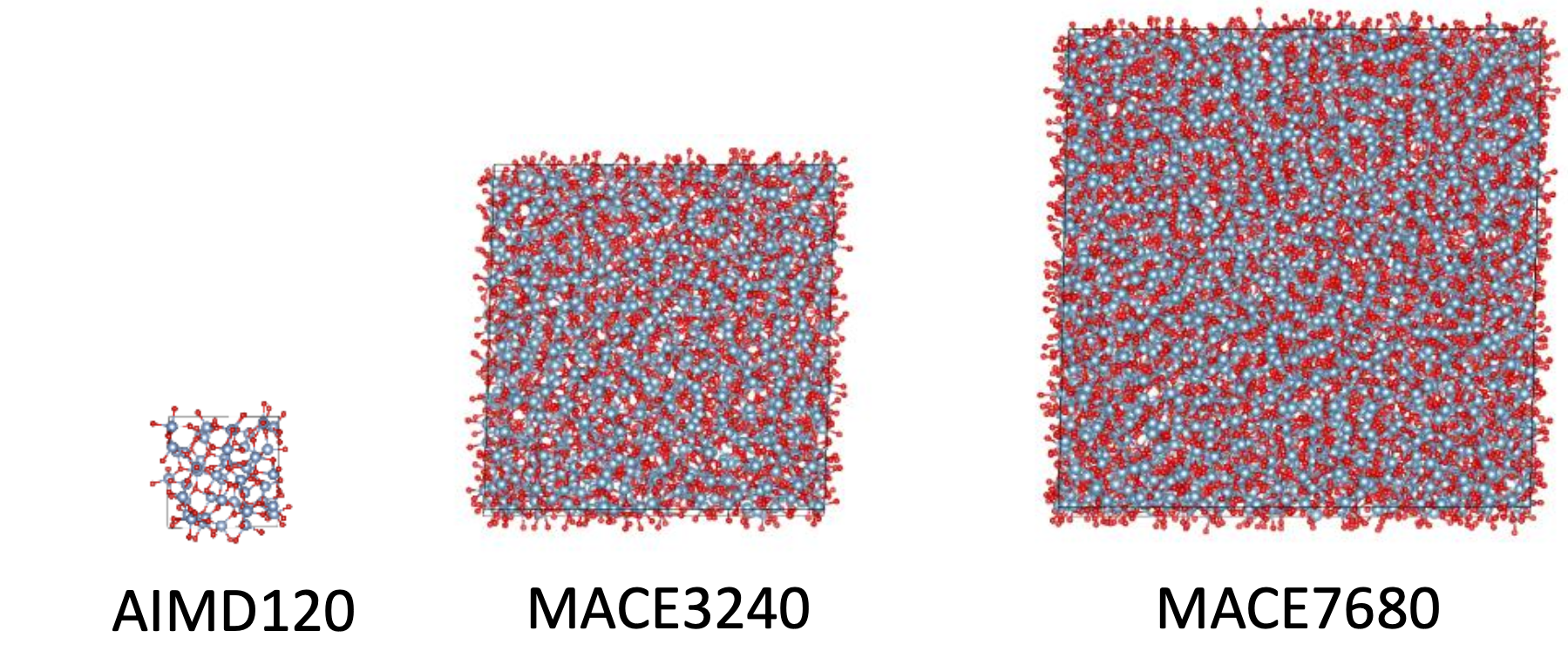}
\includegraphics[width=0.90\WidthFigure]{\folder/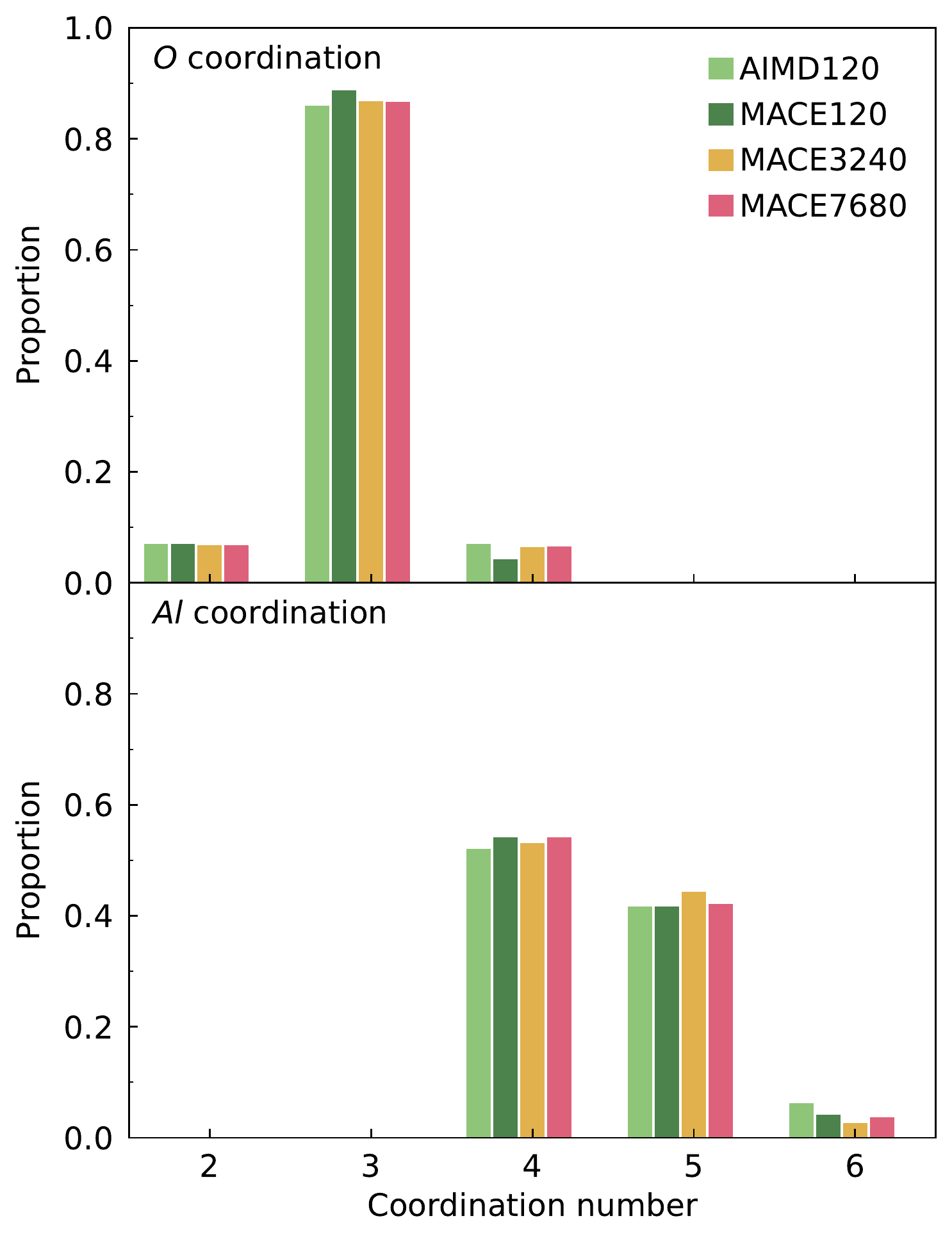}
  \caption{
  \textbf{Coordination histogram for AIMD and MACE models of am-Al$_2$O$_3$.} 
  The histograms show the proportion of oxygen (top) and aluminum (bottom) with a certain coordination number. 
  The atomistic structures above the plot show differences in the linear size of the models studied.
  }
  \label{fig:coord_hist_mace}
\end{figure}

We used MACE to generate large models of am-Al$_2$O$_3$. Specifically, we performed molecular-dynamics melt-quench simulations, using the same protocol employed in \cite{fharper_modelling_2023} to generate the AIMD 120-atom models discussed in the previous sections. 
Starting from the AIMD 120-atom model with density 2.98 g/cm$^3$ (hereafter referred to as AIMD120), we generated $3\times3\times3$ and $4\times4\times4$ supercells, containing 3240 and 7680 atoms, respectively.
These supercells were used as the initial configuration for a melt-quench simulation:
they were first heated from 0 to 4000 K in 10 ps; then melted for 10 ps at 4000 K to ensure randomness of the structure; and then quenched to 300 K over 10 ps. 
After the melt-quench simulation, each structure was equilibrated in an NVT ensemble at 300 K for 10 ps. Finally, the resulting models were relaxed to a pressure lower than $0.001$ kBar and to interatomic forces lower than 1 meV/\AA~ (without imposing any constraint on the geometry of the simulation box). 
After this final relaxation, both  the 3240- and 7680-atom models (hereafter referred to as MACE3240 and MACE7680) displayed a density of 2.92 g/cm$^3$.
We also relaxed with MACE the AIMD120 model; using the same threshold mentioned above and without any constraint on the geometry of the box, obtaining a `MACE120' model with density 2.88 g/cm$^3$.
The densities of structures obtained from first principles and MLP are in acceptable agreement, being within $3\%$.
In the next section we validate the capability of MACE to reproduce within few-percent accuracy the first-principles results for the structural, vibrational, and thermal properties of am-Al$_2$O$_3$. Then, using the 120-, 3240-, and 7680-atom MACE models, we investigate how these properties depend on the size of the model.

\subsection{Structural properties} 
\label{ssub:structural_properties}
\begin{figure}[b]
  \centering
\includegraphics[width=0.9\WidthFigure]{\folder/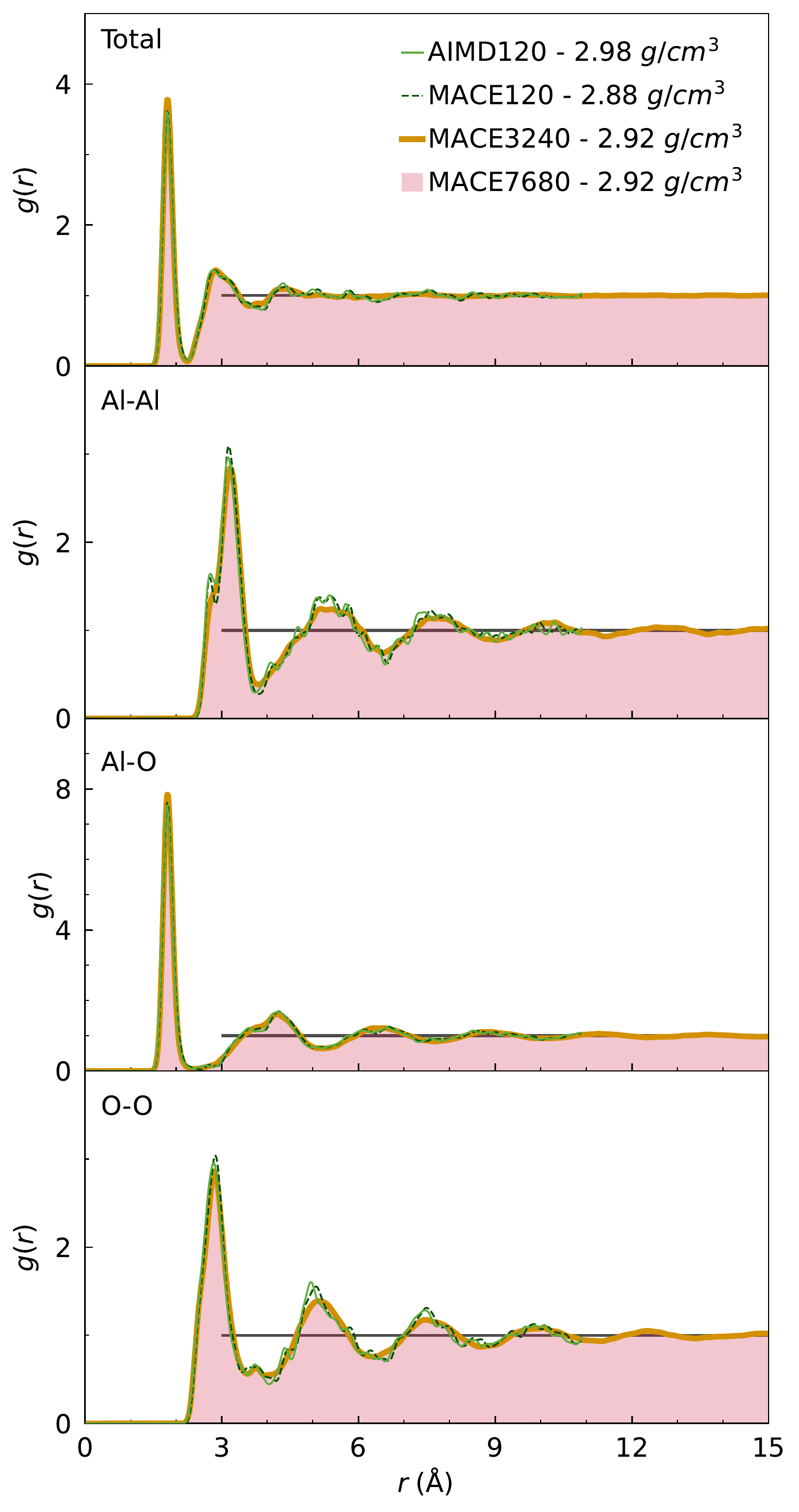}\\[-3mm]
  \caption{
  \textbf{Radial distribution function for models of am-Al$_2$O$_3$.} Total and partial RDF of the 120 atom Al$_2$O$_3$ model generated using AIMD at 2.98 g/cm$^3$ compared with the two models generated using the MACE potential.
  The RDFs for the 120-atom models only extends to $r{=}10.9$\AA, a distance equal to the linear size of these models. 
  }
  \label{fig:al2o3_rdf_mace}
\end{figure}

We start by comparing coordination histograms of MACE structures with the AIMD120 structure. Fig. \ref{fig:coord_hist_mace} shows that both Al and O coordination distributions are in good agreement between MACE120 and AIMD120. Importantly, these distributions are practically independent from the size of the model, since
MACE120, MACE3240, and MACE7680 show extremely similar distributions.

Next, to resolve possible more subtle difference between the various structures, we compare in Fig. \ref{fig:al2o3_rdf_mace} the total and partial RDFs.
The total RDF of MACE120, MACE3240, and MACE7680 are practically indistinguishable, and they are in remarkable agreement with the RDF of AIMD120.
Similarly, all the partial RDFs (Al-Al, Al-O, and O-O) are in remarkable agreement between all the four aforementioned models.
We highlight how oscillations in the partial RDFs become very weak (negligible) at distances larger than the linear size of our first-principles 120-atom models ($\sim 11$ \AA, where the solid light-green and dashed dark-green lines stop).
This suggests that atomistic models with linear size of $\sim 11$ \AA~are sufficiently large to capture the most important features of structural disorder in am-Al$_2$O$_3$.
Importantly, these tests also validate the capability of the MACE MLP to describe the structural properties of am-Al$_2$O$_3$ with first-principles accuracy. Therefore, we continue our validation tests for MACE, discussing in the next section its capability to reproduce the first-principles vibrational properties.

\subsection{Vibrational properties} 
\label{ssub:vibrational_properties}

\begin{figure}[t]
  \centering
\includegraphics[width=0.90\WidthFigure]{\folder/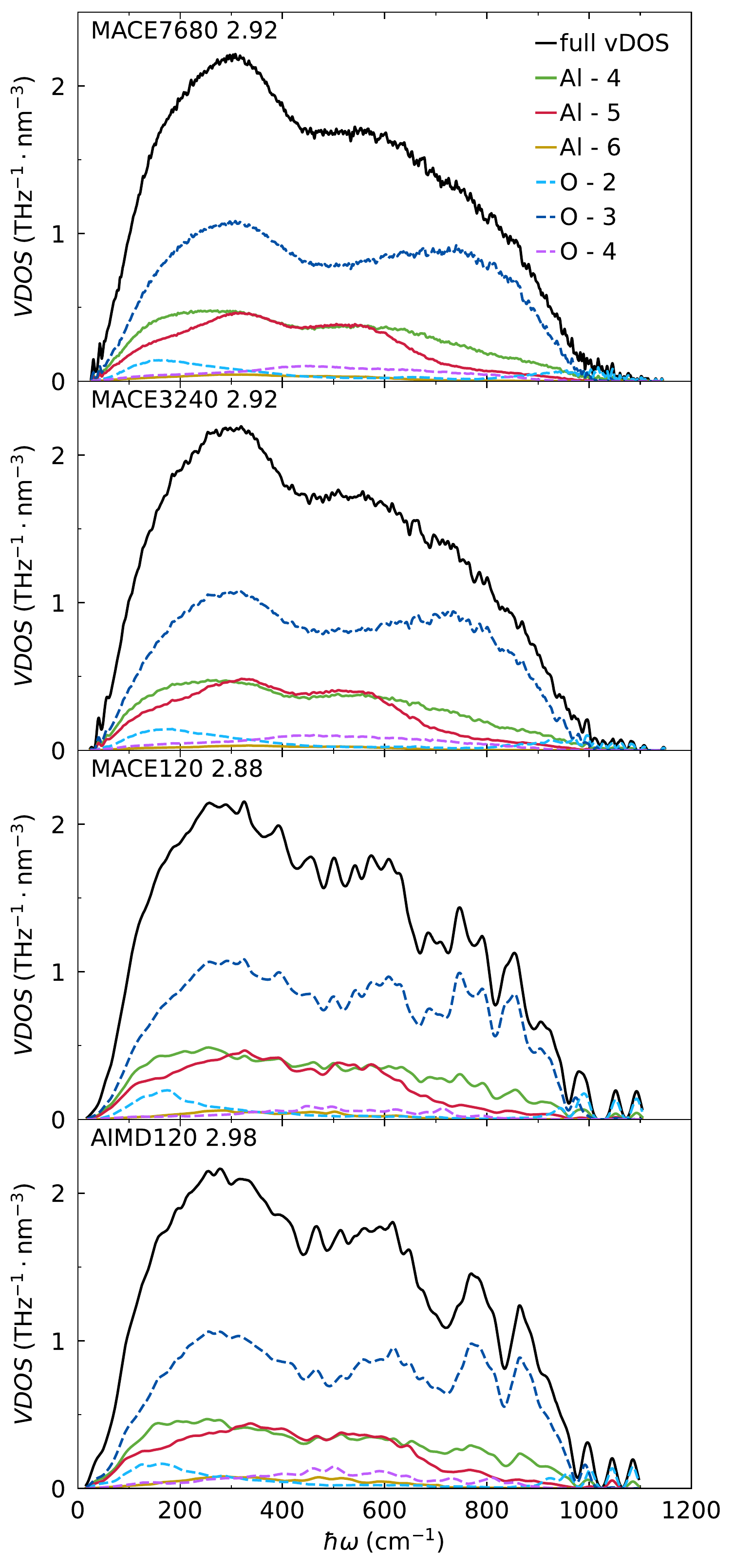}\\[-4mm]
  \caption{
  \textbf{Vibrational density of states for AIMD and MACE am-Al$_2$O$_3$ models.}
The total VDOS is solid black. The colored solid lines distinguish coordination environments for Al atoms: green is Al$_4$, red is Al$_5$, and yellow is Al$_6$. Dashed colored lines are used for coordination environments of O atoms: cyan for O$_2$, blue for O$_3$, and purple for O$_4$.
 From top to bottom:  MACE7680, MACE3240 (both computed at $\bm{q}{=}0$ only); MACE120 and AIMD120 (both computed on a 5x5x5 $\bm{q}$ mesh).
  }
  \label{fig:VDOS_mace}
\end{figure}

In Fig. \ref{fig:VDOS_mace} we test the capability of MACE to reproduce the vibrational properties previously computed from first principles, and we also check how these depend on the size of the model.
Starting from the bottom, we see that the AIMD120 and MACE120 models show very similar full vDOS (black line). Also, the decomposition of the vDOS into PDOS contributions from different coordination environments (colored lines) is compatible between MACE120 and AIMD120. 

Turning our attention to size effects on vibrational properties, we see that the large MACE3240 and MACE7680 models display a vDOS similar in magnitude and shape to the 120-atom models but differing in the roughness, \textit{i.e.,} larger models have vDOS, which is a smoothed version of the vDOS computed for small MACE120 and AIMD120 models. This is also true for the PDOS, where general shape is maintained but roughness is reduced as the model size increases.
We conclude by noting that, in all these models, the vibrational frequencies are non-negative, confirming that our models were correctly relaxed to an energy minimum and therefore are structurally stable.

\subsection{Thermal conductivity}
In this section we use the vibrational properties of the MACE models to evaluate the Wigner thermal conductivity~(\ref{eq:thermal_conductivity_combined}), and thus test how it converges with respect to system size.

\begin{figure}[b]
  \centering
\includegraphics[width=\WidthFigure]{\folder/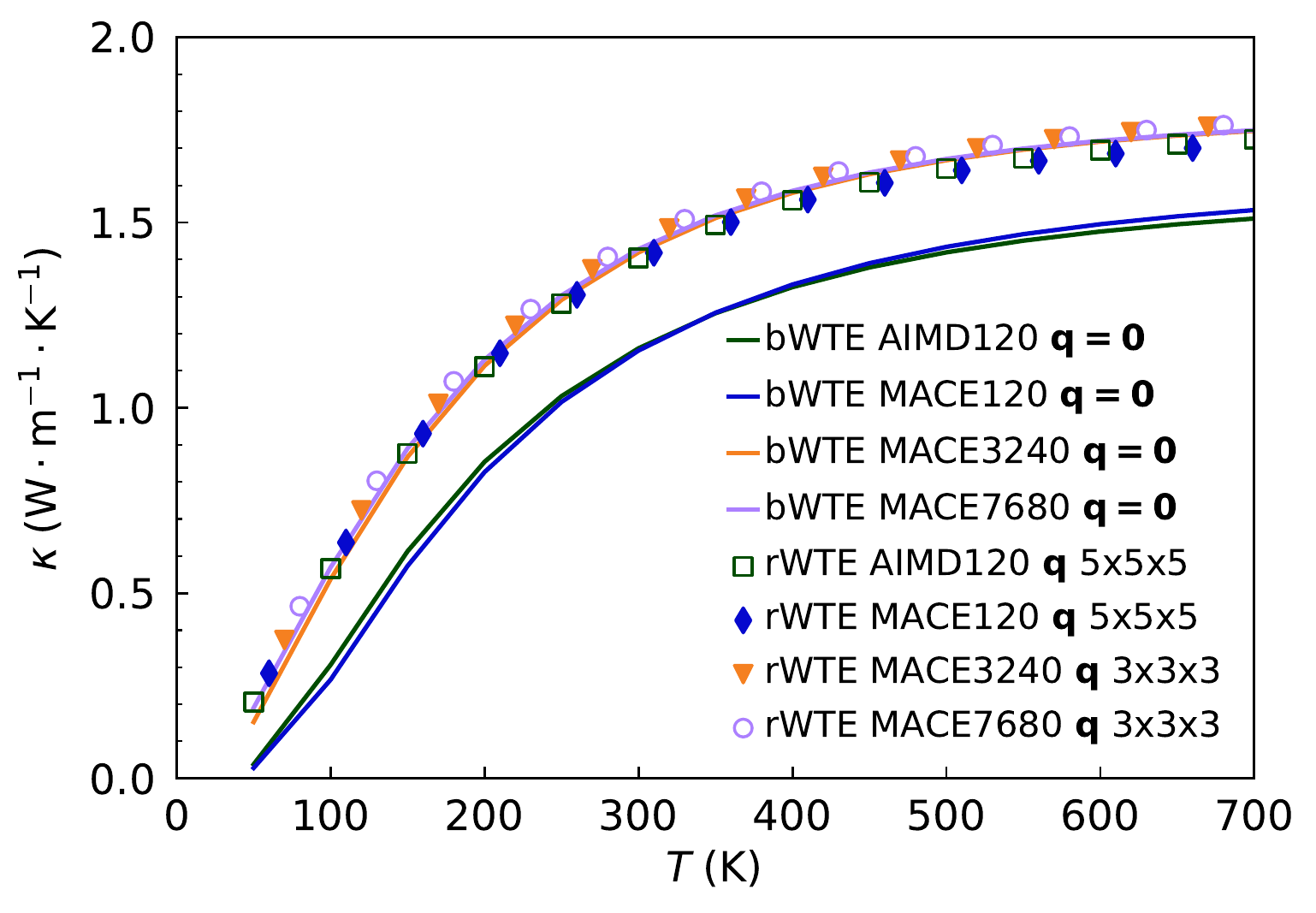}
  \caption{
  \textbf{rWTE and bWTE thermal conductivities of am-Al$_2$O$_3$ models at density $\sim$3.0 g/cm$^3$.} The solid lines (scatter points) are bWTE (rWTE) conductivities: MACE7680 is purple, MACE3240 is orange, MACE120 is blue and AIMD is green. }
  \label{fig:thermal_mace}
\end{figure}

In Fig.~\ref{fig:thermal_mace} we compare the bare WTE conductivities (\textit{i.e.}, Eq.~(\ref{eq:thermal_conductivity_combined}) computed at $\bm{q}=\bm{0}$ with $\eta=0$, hereafter referred to as `bWTE') against the regularized rWTE conductivity (computed relying on the $\bm{q}$-mesh interpolation and using $\eta$ determined from the beginning of the convergence plateau, see Figs.~\ref{fig:harm_theory_plateau},\ref{fig:mace_harm_theory_plateau} in Appendix for details).
We see that for the large MACE7680 and MACE3240 models the bWTE and rWTE conductivities are practically indistinguishable, confirming that these models are sufficiently large to realistically represent a bulk system at temperatures above 50 K, thus the convergence-acceleration protocol based on the Voigt regularization protocol has negligible effect on them.
Importantly, the rWTE conductivity for the small AIMD120 and MACE120 models computed on converged 5x5x5 $\bm{q}$-mesh and using $\eta=6$ cm$^{-1}$ yields a conductivity compatible \footnote{The largest difference between the converged conductivities is of about 3.3$\%$ ($\approx$ 0.057 W/mK) observed at 700 K between the rWTE computed over a 5x5x5 $\bm{q}$-mesh for MACE120, and rWTE computed over a 3x3x3 $\bm{q}$-mesh for MACE7680.} with that of the large MACE7680 and MACE3240 models.
We highlight that it is crucial to employ the rWTE to extrapolate the bulk limit of the conductivity from small models---Fig.~\ref{fig:thermal_mace} shows that in AIMD120 and MACE120 the bWTE underestimates the conductivity by approximately $20\%$.

It is worth commenting on how the rWTE protocol extrapolates successfully to the bulk limit for the conductivity.
As anticipated in Sec.~\ref{ssec:Wigner_formulation} and in Ref.~\cite{simoncelli_thermal_2022}, the rWTE protocol
enforces the physical property that couplings between different modes can always occur in a truly disordered system, and exploits $\bm{q}$-mesh interpolation
to average the vibrational properties over many possible different boundary conditions---this last operation is analogous to the averaging over periodic and anti-periodic boundary conditions employed by \citet{feldman_thermal_1993} to accelerate computational convergence.
To better understand the effect of the $\bm{q}$-mesh interpolation, we show in Fig.~\ref{fig:vel_op_mace_2D}
that the velocity-operator as a function of frequency for MACE120 on a $5\times5\times 5$ $\bm{q}$ mesh, approaches the velocity operator as a function of frequency computed at $\bm{q}=\bm{0}$ for the larger MACE7680 model.
We note that to represent the velocity operator as a function of frequency using Eq.~(\ref{eq:v_operator_omega_a_omega_d}) in a finite-size model, the Dirac delta must be broadened with a finite-width (Gaussian) distribution having width of the order of few energy-level spacings $\hbar\Delta\omega_{\rm avg}$, \textit{i.e.} a broadening comparable to the value $\eta$ used in the evaluation of the AF conductivity.
In the context of Eq.~(\ref{eq:thermal_conductivity_combined}), employing a broadening of the order of $\hbar\Delta\omega_{\rm avg}$ ensures in practice that such an expression behaves analogously to that describing a strongly disordered system in the bulk limit. The vibrational eigenstates repel each other significantly \cite{simkin_minimum_2000}; therefore, vibrational modes are dense but never degenerate and, for arbitrarily small values for the broadening $\eta$, couplings between neighboring vibrational eigenstates are allowed.
The model size necessary to achieve computational convergence with the rWTE protocol depends on the degree of disorder present: for strongly disordered systems such as am-Al$_2$O$_3$ and vitreous silica \cite{simoncelli_thermal_2022,ndour_practical_2023}, models containing hundreds of atoms are sufficient to reproduce the bulk limit; future work will investigate systems with lower degree of disorder.

\begin{figure}[b]
  \centering
\includegraphics[width=\WidthFigure]{\folder/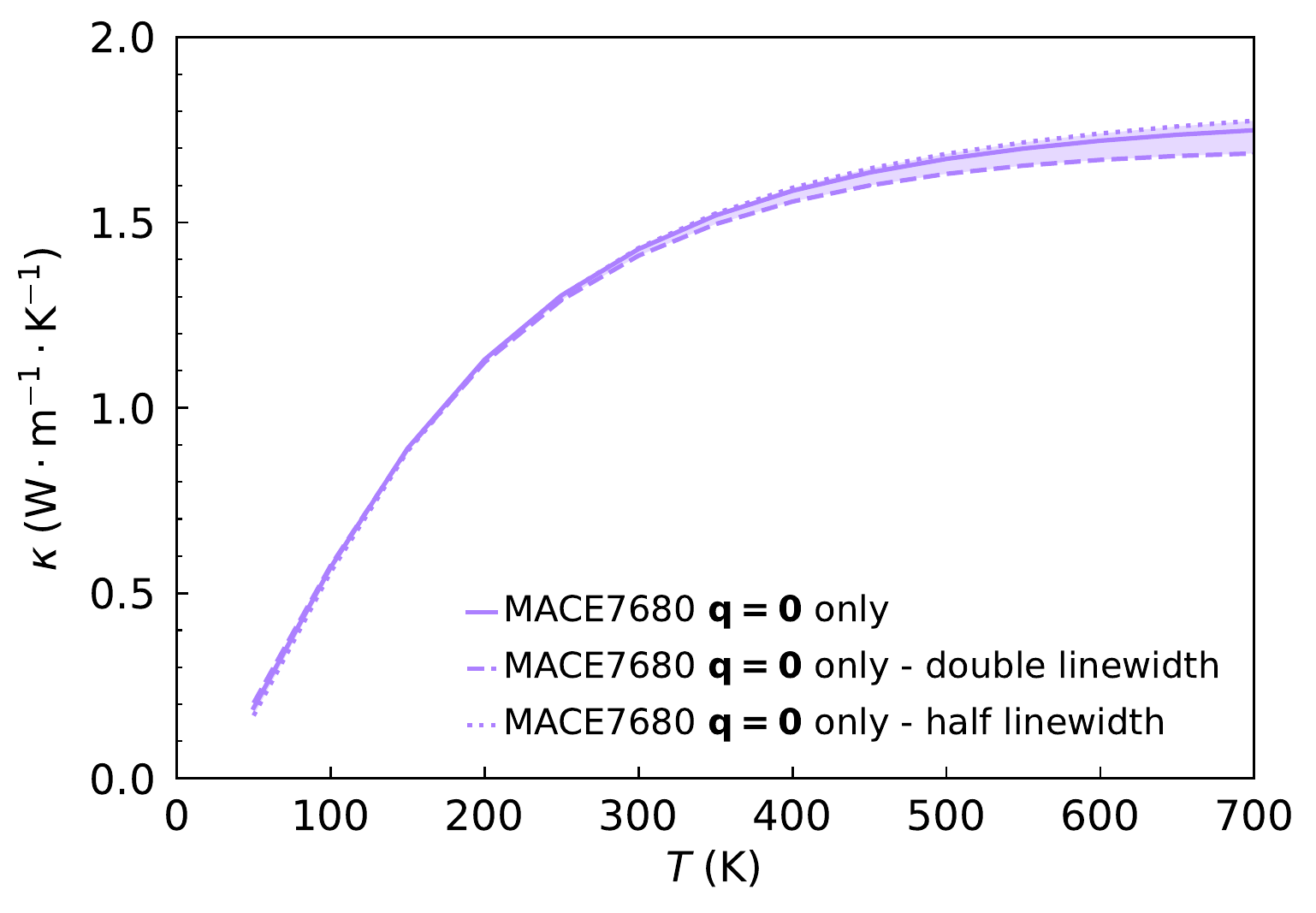}
  \caption{
  \textbf{Negligible effect of anharmonicity on thermal conductivity of am-Al$_2$O$_3$.} Solid, bWTE conductivity of MACE7680, calculated with physical linewidths. The dashed and dotted lines are bWTE conductivities calculated by artificially doubling and halving the anharmonic linewidths, respectively. We see that these artificial rescalings negligibly affect the conductivity, indicating that structural disorder dominates over anharmonicity in determining heat transfer in am-Al$_2$O$_3$.
  }
  \label{fig:thermal_mace_anharmonicity}
\end{figure}

To further validate our assertion that structural disorder is the dominant limiting factor for heat conduction in am-Al$_2$O$_3$ over the range 50-700 K, we selected the MACE7680 model---which is large enough to directly compute the bulk conductivity, without regularization---and 
calculated the bWTE conductivity in three ways:
(i) with `physical' linewidths computed from first principles using the AIMD120 model (solid line in Fig.~\ref{fig:thermal_mace_anharmonicity}); 
(ii) using artificially enlarged linewidths, obtained multiplying by a factor 2.0 the physical linewidths (dashed line); (iii) using artificially reduced linewidths, obtained multiplying the physical ones by a factor 0.5 (dotted line). 
Intuitively, when structural disorder dominates over anharmonicity in determining heat conduction, we expect the conductivity to be practically indistinguishable in the three cases above. This behavior is in sharp contrast with that observed in ordered simple crystals with well separated phonon bands---in this case anharmonicity determines the conductivity, and in the high-temperature limit doubling the linewidth directly implies a reduction of the conductivity by a factor of two (one can verify this analytically by simply rescaling the linewidths appearing in the thermal conductivity expression for crystals, e.g. Eq.~(49) of Ref.~\cite{simoncelli2021Wigner}). Fig.~\ref{fig:thermal_mace_anharmonicity} shows that artificially rescaling the linewidths produces practically unnoticeable changes in the thermal conductivity of am-Al$_2$O$_3$ between 50 and 700 K, showing that in this system structural disorder dominates over anharmonicity in determining heat conduction. 
In other words, keeping in mind that Eq.~(\ref{eq:thermal_conductivity_combined}) shows that heat conduction in amorphous solids is mediated by couplings between vibrational modes, Fig.~\ref{fig:thermal_mace_anharmonicity} shows that in strongly disordered solids such as  am-Al$_2$O$_3$ anharmonicity only serves to allow couplings between modes, and as soon as anharmonicity is large enough (in the thermodynamic limit an infinitesimal value is sufficient), its exact magnitude is unimportant and does not influence the value of the conductivity. This behavior is related to the presence of velocity-operator elements between pairs of vibrational eigenstates that do not significantly depend on the energy difference between the eigenstates coupled. 
These findings are consistent with the analysis reported in Fig. \ref{fig:vel_op_all_2D}, as well as with the negligible differences observed between the  AF and rWTE conductivities between 50 and 700 K.

\section{Conclusions}

The ubiquitous use of am-Al$_2$O$_3$ in electronic devices and the several open fundamental questions on how its atomistic structural and vibrational properties determine its macroscopic thermal properties prompted us to study this material from a first principles level of theory. 
We generated and characterized atomistic models of am-Al$_2$O$_3$ from AIMD with densities ranging from 2.28 g/cm$^3$ to 3.49 g/cm$^3$, describing how the atomic coordination topology varies with density. 
We have shown that at least five different atomic coordination environments coexist in am-Al$_2$O$_3$, and these lead to significant structural disorder already at the sub-nanometre lengthscale. 
We have discussed how the atomic coordination topology affects the vibrational properties, showing that
different coordination environments for oxygen and aluminium have fingerprints on the coordination-resolved PDOS.
We have described the thermal properties using the recently introduced Voigt-regularized Wigner formulation (rWTE) \cite{simoncelli_thermal_2022}, accounting comprehensively for the effects of structural disorder, anharmonicity, and quantum Bose-Einstein statistics. 
We have shed light on the microscopic physics underlying thermal transport in am-Al$_2$O$_3$, discussing the dominant role played by strong structural disorder (emerging from having at least five coexisting different atomic coordination topologies) and the negligible role played by anharmonicity.
Specifically, we showed that the harmonic Allen-Feldman theory---evaluated using the convergence-acceleration protocol discussed in Ref.~\cite{simoncelli_thermal_2022}---yields predictions in close agreement with the anharmonic rWTE protocol and with experiments.
We have validated these first-principles calculations using a MACE MLP, generating models of am-Al$_2$O$_3$ containing 3240 and 7680 atoms at density $\sim 3$ g/cm$^3$, showing that their thermal conductivity is compatible with that of the 120-atom first-principles models. 
We discussed how the increase in the thermal conductivity observed with density derives from an increase of the vibrational density of states with density, as well as from an increase of the diffusivity with density.
Importantly, we have investigated the thermal properties also below room temperature ($T\gtrsim 50 K$), where the quantum Bose-Einstein statistics of vibrations yields a specific heat significantly different from the classical limit, providing information on the thermal conductivity in a regime inaccessible by molecular-dynamics-based methods \cite{li_effects_2020,PhysRevMaterials.3.085401}, which are governed by classical equipartition \cite{PhysRevMaterials.3.085401} and thus limited to high temperatures. 
Ultimately, this study further validates the capability of the recently developed rWTE protocol \cite{simoncelli_thermal_2022} to describe the thermal properties of strongly disordered glasses using atomistic models containing hundreds of atoms, and thus within the reach of first-principles techniques.

\section*{Acknowledgements}
A. F. H. acknowledges the financial support of the Gates Cambridge Trust and the Winton Programme for the Physics of Sustainability, University of Cambridge.
K.I. acknowledges support from Winton \& Cavendish Scholarship at the Department of Physics, University of Cambridge.
W.C.W. and M.C.P. acknowledge support from the EPSRC (Grant EP/V062654/1).
M. S. acknowledges support from Gonville and Caius College, and from the SNSF project P500PT\_203178. The calculations presented in this work have been performed using computational resources provided by: (i) the Sulis Tier 2 HPC platform hosted by the Scientific Computing Research Technology Platform at the University of Warwick (Sulis is funded by EPSRC Grant EP/T022108/1 and the HPC Midlands+ consortium); (ii) Cambridge Tier-2 system operated by the University of Cambridge Research Computing Service (www.hpc.cam.ac.uk) funded by EPSRC Tier-2 capital grant EP/T022159/1, and GPU resources were obtained through a University of Cambridge EPSRC Core Equipment Award, EP/X034712/1.

\newpage
\appendix

\section{Computational details} 
\label{sec:computational_details}

\subsection{First-principles calculations} 
\label{sec:computational_details}
The 120-atom am-Al$_2$O$_3$ structures were obtained from the open-source dataset of Ref.~\cite{harper2023research}. These models were generated via first-principles molecular dynamics simulations, using the melt quench procedure described in Ref.~\cite{fharper_modelling_2023}, with VASP v5.4 \cite{kresse1993ab}.
In order to apply the computational protocol of Ref.~\cite{simoncelli_thermal_2022} to study the thermal properties, the vibrational frequencies have to be interpolated in Fourier space. 
Applying Fourier interpolation to the vibrational frequencies of disordered atomistic models containing less than 200 atoms yields more accurate results when a mesh denser than the point $\bm{q}=\bm{0}$ only is used as starting point. 
Therefore, to compute the vibrational properties of am-Al$_2$O$_3$ on a mesh denser than $\bm{q}=\bm{0}$ only and to use the most accurate density-functional perturbation theory (DFPT) technique \cite{baroni_phonons_2001}, the vibrational properties are computed using Quantum ESPRESSO \cite{giannozzi_quantum_2009,giannozzi_advanced_2017} on a $2\times2\times2$ $\bm{q}$-mesh (at present the DFPT implementation in VASP is restricted to calculations at $\bm{q}{=}\bm{0}$). Quantum ESPRESSO calculations were carried out using PBE functional, with pseudopotentials from the standard solid-state pseudopotential libraries (SSSP) precision library \cite{prandini2018precision}.
To compute the vibrational properties, the cell and atomic positions of all the am-Al$_2$O$_3$ models were relaxed using a threshold for forces of 2$\times$10$^{-4}$\,Ry/Bohr and of $0.01$ kBar for pressure (\texttt{vc-relax} command).

\subsection{Generation of the MACE potential} 
\label{sub:fit_of_the_mace_machine_learning_potential}
The MACE MLP was trained using the first-principles (PBE functional) dataset from \cite{li_effects_2020}.
We used a two-layer MACE and a per-layer cutoff of 4.5 \AA, resulting in a total receptive field of 9 \AA, as well as 128 embedding channels and $L_{\mathbf{max}}=1$. Both energies and forces were used in training and ten percent of the full dataset was randomly held out for validation; the remaining ninety percent comprised the training set.
The training proceeds until the onset of overfitting is observed: we monitor the errors on the validation set during training and exit when those errors begin to increase.
Following this procedure, the training terminated with a validation error of 5.4 meV/atom for the energies and 123.1 meV/\AA{}  for the forces. 

\section{Coordination environment} 
\label{sec:coord_rdf_diff}
{To determine the coordination topology we calculated the number of atoms in a sphere of radius equal to the first minimum of the radial distribution function (Fig.~\ref{fig:al2o3_rdf}), which ranged from e.g., 2.2 \AA \space for the lowest-density structure, to  2.35 \AA \space for the highest-density structure. }

\begin{figure*}
  \centering
\includegraphics[width=\textwidth]{\folder/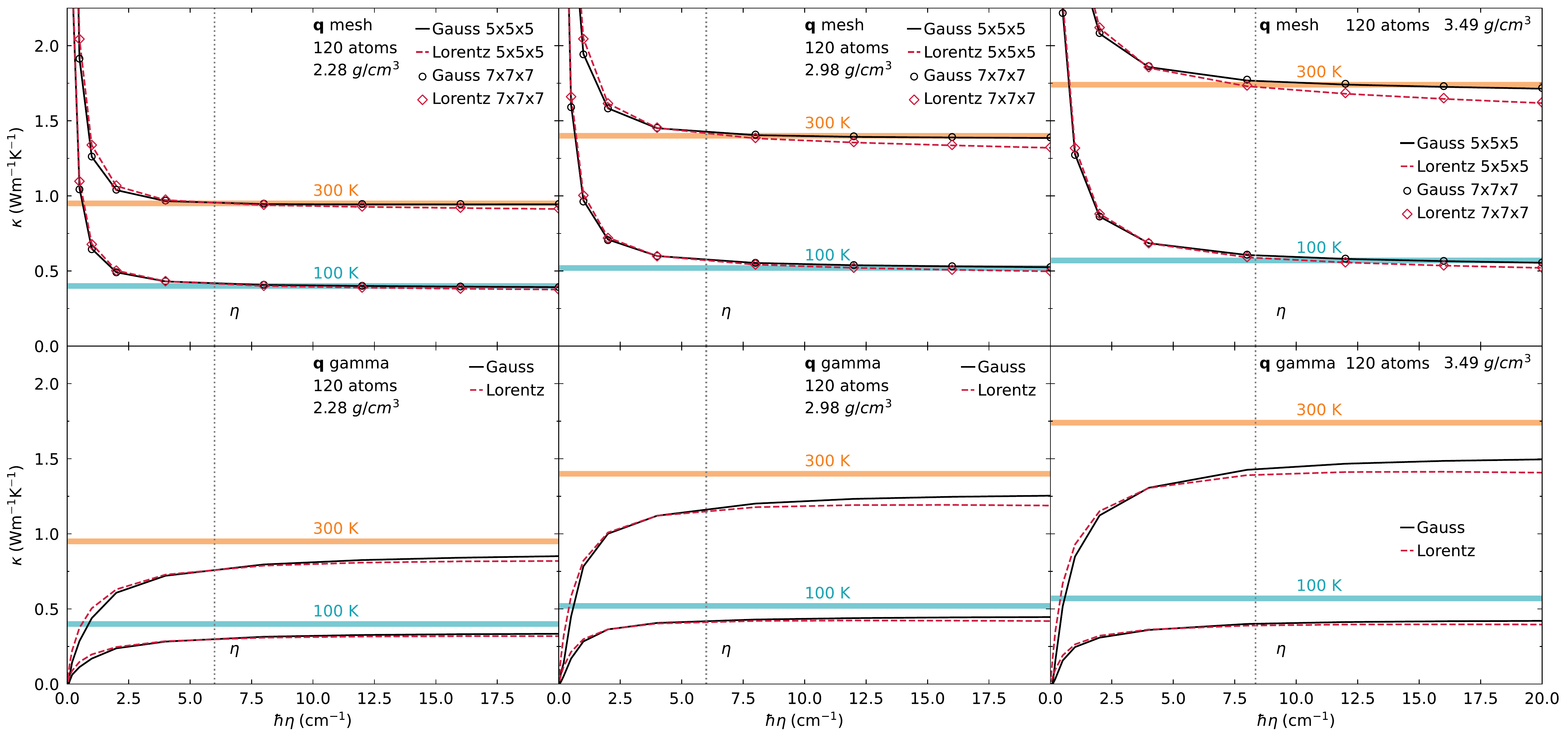}\\[-3mm]
  \caption{\textbf{Convergence of the AF conductivity for am-Al$_2$O$_3$ with respect to the broadening $\eta$ for the Dirac $\delta$,} The three upper panels show convergence plateaus for $5\times5\times5$ (lines) and $7\times7\times7$ (scatter points) meshes of models with densities 2.28, 2.98 and 3.49 $g / cm^3$. Black (red) correspond to evaluating the AF conductivity using the Gaussian (Lorentzian) representation of the Dirac delta function. 
  The three lower panels contain calculations at $\bm{q}=\bm{0}$ only for models with densities 2.28, 2.98 and 3.49 $g / cm^3$.  Black and red denote Gaussian and Lorentzian, as in the panels above. The broadening $\eta$ is chosen as the value approximately determining the beginning of the convergence plateau \cite{simoncelli_thermal_2022}, and it is set to 6.0 cm$^{-1}$ for models with density up to 3.30 $g / cm^3$ and 8.34 for the model with density $3.49 g/cm^3$. We see that the $5\times5\times5$ $\bm{q}$ mesh is dense enough to achieve computational convergence with respect to mesh size, since results obtained on a $7\times7\times7$ mesh are practically indistinguishable from those obtained using a $5\times5\times5$ mesh. Calculations at $\bm{q}=\bm{0}$  only are far from computational convergence and consequently underestimate the conductivity. 
}
  \label{fig:harm_theory_plateau}
\end{figure*}
\section{First-principles thermal conductivity calculations}
\subsection{Convergence of the Allen-Feldman theory} 
\label{sec:convergence_of_the_allen}

In order to calculate the bulk limit of the thermal conductivity of strongly disordered solids such as am-Al$_2$O$_3$, we rely on the convergence-acceleration protocol discussed in Ref.~\cite{simoncelli_thermal_2022} both for the AF and rWTE conductivities. The capability of such a protocol to accurately extrapolate the bulk limit of the thermal conductivity of strongly disordered glasses from finite-size models containing hundreds of atoms is  validated in Sec.~\ref{sec:size_effects}, and in Ref.~\cite{simoncelli_thermal_2022}.
{The protocol requires determining the broadening parameters $\eta$ for the Voigt profile appearing in Eq.~(\ref{eq:thermal_conductivity_combined}) as a value determining the beginning of the convergence plateau shown in Fig.~\ref{fig:harm_theory_plateau} (see Sec.~\ref{ssec:Wigner_formulation} for details). 

All the am-Al$_2$O$_3$ models analyzed display a clear and broad convergence plateau for the AF conductivity.
The three upper panels in Fig. \ref{fig:harm_theory_plateau} show results obtained employing a $5\times5\times5$ or $7\times7\times7$ $\bm{q}$-mesh; the good agreement between these two calculations indicates that computational convergence has been achieved.
 
The three bottom panels of Fig.~\ref{fig:harm_theory_plateau} show that a calculation performed at $\bm{q}=0$ only for a 120-atom model is far from computational convergence and underestimates the thermal conductivity.
The values of $\eta$ that we determined from the convergence test discussed in Fig. \ref{fig:harm_theory_plateau} and that we employed in our calculations are reported in Table \ref{tab:broadening}.

\begin{table}[h]
    \centering
    \begin{tabular}{|c|c|c|c|c|c|}
    \hline
    $\rho$ ($g/cm^3$) & 2.28 & 2.98 & 3.17 & 3.30 & 3.49 \\
    \hline
    $\hbar\eta$ (cm$^{-1}$) & 6.0 & 6.0 & 6.0 & 6.0 & 8.34 \\
    \hline
    \end{tabular}
    \caption{Broadening parameters $\eta$ used for the Gaussian representation of the Dirac $\delta$  function appearing in the AF conductivity expression, and for the Voigt distribution appearing in the rWTE expression.}
    \label{tab:broadening}
\end{table}

\subsection{Radial Distribution Functions} 
\label{sub:radial_distribution_functions}
All the radial distribution functions are computed following Ref.~\cite{LEROUX201070}, using a Gaussian distribution with standard deviation equal to 0.075 \AA.

\subsection{Anharmonic Linewidths} 
\label{sec:linewidths}

\begin{figure*}
  \centering
\includegraphics[width=\textwidth]{\folder/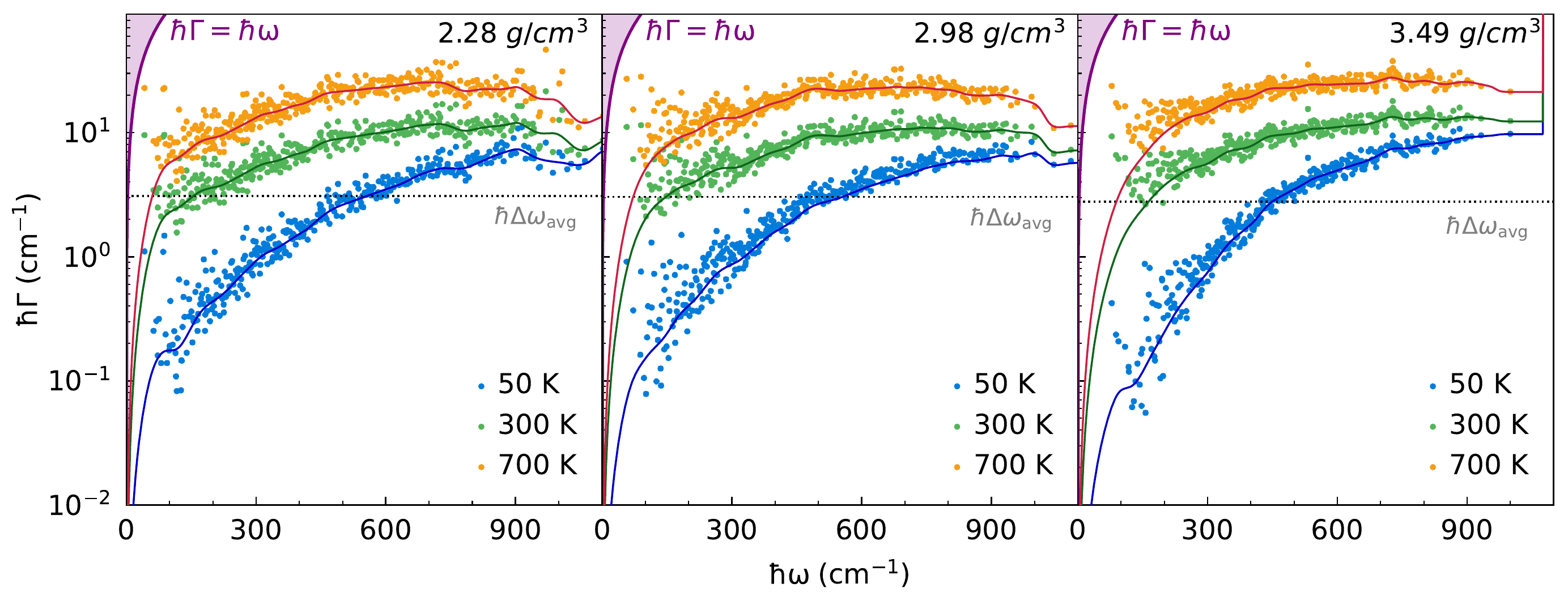}\\[-3mm]
  \caption{\textbf{Effect of temperature on anharmonic linewidths of am-Al$_2$O$_3$ at various densities and temperatures.} The scatter points represent the linewidths computed at $\bm{q}=\bm{0}$, and the solid lines are coarse grained functions $\Gamma_a[\omega, T]$ used to approximately describe the anharmonic linewidths as a single-valued functions of frequency, thus to estimate the effects of anharmonicity at a reduced computational cost\cite{PhysRevB.105.134202,PhysRevLett.106.045901,simoncelli2021Wigner}. The purple region denotes the overdamped regime $\Gamma > \omega$ \cite{simoncelli2021Wigner,caldarelli_many-body_2022}. The gray dashed lines show the average spacing between vibrational energy levels. We note that the linewidths of am-Al$_2$O$_3$ are similar to those found other oxide glasses, e.g., vitreous silica \cite{simoncelli_thermal_2022}. 
  }
  \label{fig:temp_v_linewidth}
\end{figure*}

The third-order interatomic force constants are computed in the 120 atom cells using ShengBTE \cite{li2014shengbte} up to the 8$^{\mathrm{th}}$ nearest neighbor. The linewidths were then computed using \texttt{phono3py} \cite{phono3py,togo_first-principles_2023}, with a Gaussian smearing of 0.18 THz or \textasciitilde 6 cm$^{-1}$, and the standard perturbative treatment of anharmonicity, \textit{i.e.} (i) vibrational frequencies were considered to be independent from temperature (\textit{i.e.} it neglects thermal expansion and the renormalization of frequencies due to anharmonicity \cite{monacelli_stochastic_2021,tadano_first-principles_2022,jain_multichannel_2020,feng_four-phonon_2017,PhysRevMaterials.3.085401}); (ii) the linewidths were computed considering exclusively the cubic terms in the Taylor expansion of the interatomic potential \cite{simoncelli2021Wigner,simoncelli_thermal_2022} and the contribution due to isotopic-mass disorder \cite{tamura_isotope_1983}.

Fig. \ref{fig:temp_v_linewidth} shows the linewidths for AIMD structures calculated at $\bm{q}=\bm{0}$ at various densities and temperatures. {We see that at 50K a significant proportion of vibrational modes have linewidths smaller than the average energy-level spacing. As mentioned in section \ref{sec:thermal_properties}, these linewidths are employed within the Voigt distribution, which ensures that heat transfer between neighboring vibrational eigenstates can always occur, implying that the effects of anharmonicity are accounted for only when they are not altered by finite-size effects \cite{simoncelli_thermal_2022}.

The solid lines in Fig. \ref{fig:temp_v_linewidth} 
are the  functions $\Gamma_a[\omega]$ that approximately describe the anharmonic linewidths as a single-valued functions of frequency, obtained 
following the approaches discussed in Ref.~\cite{simoncelli2021Wigner} (see also Refs.~\cite{PhysRevB.105.134202,PhysRevLett.106.045901,Fiorentino_2023} for similar approximated treatments of anharmonicity).
The approximated function $\Gamma_a[\omega]$ is employed to compute the anharmonic linewidths as a function of frequency when the Fourier interpolation is used to extrapolate the bulk limit of the thermal conductivity~\ref{eq:thermal_conductivity_combined}, following the protocol discussed in Ref.~\cite{simoncelli2021Wigner}.

\subsection{rWTE conductivity calculation}
The quantities needed to evaluate the rWTE conductivity~(\ref{eq:thermal_conductivity_combined}) were computed as follows:
(i) the $\eta$ parameter was computed as discussed in Sec.~\ref{sec:convergence_of_the_allen}; 
(ii) the software \texttt{phono3py} \cite{phono3py,togo_first-principles_2023} was used to evaluate frequencies and velocity operators on a $\bm{q}$ mesh;
(iii) the linewidths were evaluated from the frequencies determined at the previous point using the function $\Gamma_a(\omega)$ discussed in Sec.~\ref{sec:linewidths}.

We checked that increasing the size of the $\bm{q}$ mesh from $5\times5\times5$ to $7\times7\times7$ produced practically indistinguishable results in the RMDS, VDOS, and thermal conductivities. Therefore, all the results discussed in the main text are evaluated on a $5\times5\times5$ $\bm{q}$ mesh.

We limited our calculations to temperatures higher than 50 K, since at temperatures lower than 50 K the thermal properties of am-Al$_2$O$_3$ are dominated by low-frequency vibrational modes that are likely to feature glassy anomalies \cite{schirmacher2006thermal,lubchenko2003origin,wang_low-frequency_2019}; accurately sampling these low-frequencies vibrational modes requires using atomistic models containing thousands of atoms, and it is therefore beyond the scope of the present work.

{$D(\omega)$ data for the plot in Fig. \ref{fig:diffusivity} was calculated using Eq.~(\ref{eq:diff_omega}) with $\delta$-function smeared to a Gaussian with variance $\frac{\pi}{2} \eta_0^2$, with $\hbar\eta_0 = 8.34$ cm$^{-1}$.}

\section{Thermal conductivity calculations using MACE} 
\label{sec:thermal_conductivity_calculations_using_mace}
\begin{figure*}
  \centering
\includegraphics[width=\textwidth]{\folder/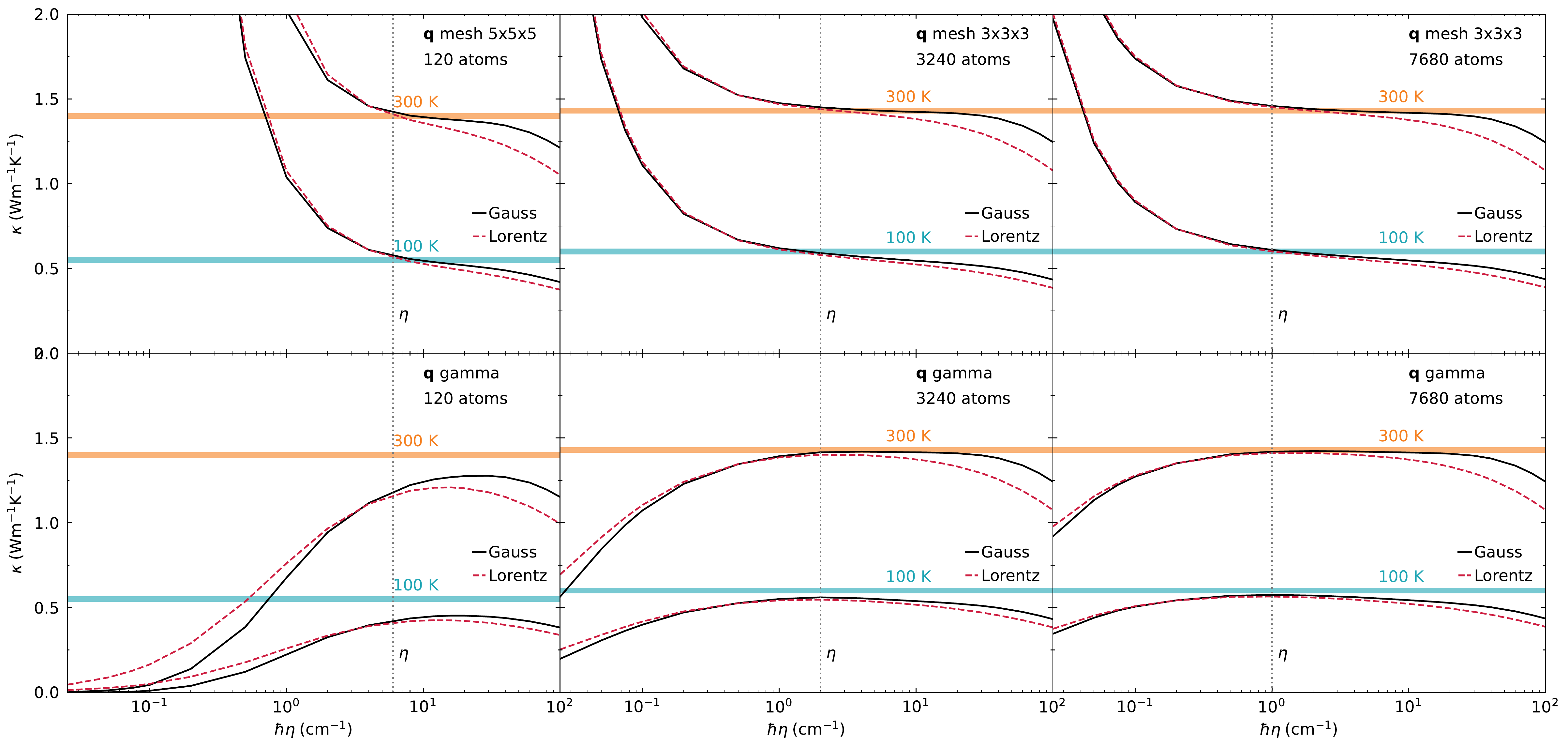}\\[-3mm]
  \caption{\textbf{Convergence of the AF conductivity for MACE models of am-Al$_2$O$_3$ with respect to the broadening $\eta$ for the Dirac $\delta$}, The three upper panels show convergence for 5x5x5 and 3x3x3 meshes of MACE models containing 120, 3240 and 7680 atoms. Black (red) correspond to evaluating the AF conductivity using the Gaussian (Lorentzian) representation of the Dirac delta. The three lower panels contain calculations at $\bm{q}{=}0$ for MACE models with 120, 3240 and 7680 atoms. Black and red denote Gaussian and Lorentzian, as in the panels above. The broadening $\eta$ is chosen as the value approximately determining the beginning of the convergence plateau \cite{simoncelli_thermal_2022}, and it is set to 6.0 cm$^{-1}$ for MACE120, 2.0 cm$^{-1}$ for MACE3240, and 1.0 cm$^{-1}$ for MACE7680; the values of $\eta$ are shown as dotted vertical black lines. We see that a model size of 3240 atoms is sufficient to obtain equivalent results of AF thermal conductivity using the $\bm{q}$-mesh or $\bm{q}{=}0$ point only, and using larger models extends the plateau to lower values of $\eta$, as in vitreous silica \cite{simoncelli_thermal_2022}. 
}
  \label{fig:mace_harm_theory_plateau}
\end{figure*}

For the MACE models, we present their convergence tests in Fig. \ref{fig:mace_harm_theory_plateau}. All three structures display a broad convergence plateau for AF conductivity on a $\bm{q}$-mesh that is increased in length as one increases the system size. We further note that 3240-atom and 7680-atom structures are both big enough to exhibit a convergence plateau at $\bm{q}{=}0$ point only that is compatible with the plateau at 3x3x3 $\bm{q}$-mesh indicating achievement of computational convergence.

Given that Fig.~\ref{fig:thermal_mace_anharmonicity} shows that anharmonicity has negligible effects on the conductivity of am-Al$_2$O$_3$, in the calculation of bWTE and rWTE conductivity for 120, 3240 and 7680-atom MACE models we used the coarse-grained functions for the anharmonic linewidths derived from  the 2.98 g/cm$^3$ AIMD model (central panel in Fig.~\ref{fig:temp_v_linewidth}).

\newpage
\section{Velocity operator for MACE and AIMD models at densities close to 3.0 g/cm$^3$}

To compare predictions of AIMD and MACE  for the velocity operator, we plot in Fig. \ref{fig:vel_op_mace_2D} the  velocity operator represented as a function of frequency difference and average ($\langle | \nu^{\mathrm{avg}}_{\omega_a\omega_d} |^2 \rangle$, Eq.~\ref{eq:v_operator_omega_a_omega_d}). In the upper panel, we can see comparison between AIMD and MACE structures with 120 atoms each. We note the agreement between different methods is  satisfactory, especially between 200 and 700 cm$^{-1}$, where vibrational DOS is at its maximum.
In the middle panel of Fig. \ref{fig:vel_op_mace_2D} we compare $\bm{q}{=}\bm{0}$ and 5x5x5 $\bm{q}$-mesh calculations of velocity operator elements for 120-atom MACE model. We see that $\bm{q}{=}\bm{0}$ calculation underestimates velocity operator elements for frequencies below 150 cm$^{-1}$, ultimately leading to underestimate of thermal conductivity as discussed in Fig. \ref{fig:thermal_mace}.
In the lower panel of Fig. \ref{fig:vel_op_mace_2D} we compare the velocity operator elements for a small 120-atom model averaged over a $\bm{q}$-mesh, with those at $\bm{q}{=}\bm{0}$ for a large 7680-atom model: these are overall similar, explaining the compatible predictions for rWTE conductivities shown in Fig. \ref{fig:thermal_mace}, and validating the idea of obtaining velocity operator elements averaging over different boundary conditions in small models of disordered solids.

\begin{figure}[t!]
    \centering
    \includegraphics[width=\WidthFigure]{\folder/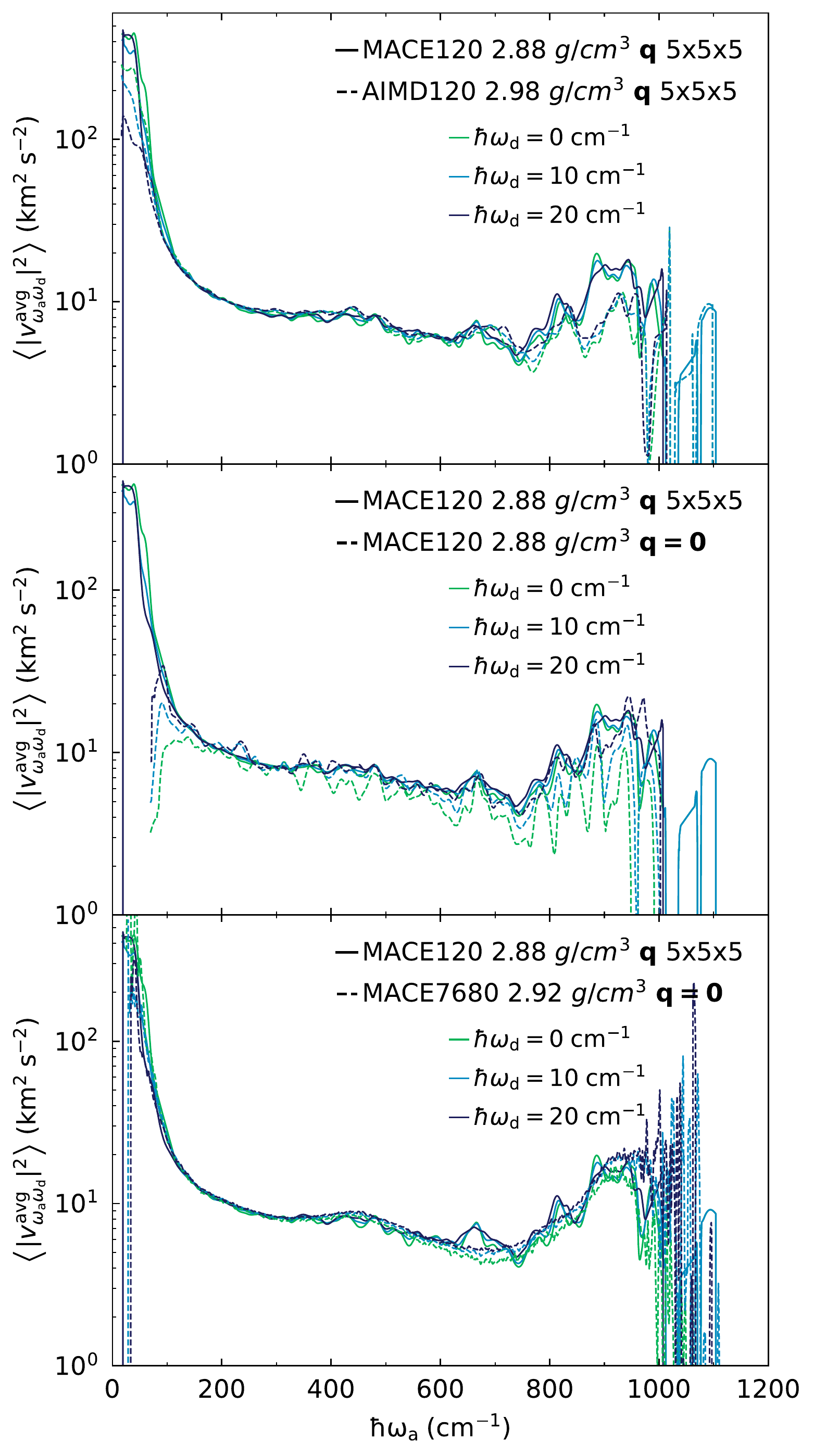}
    \caption{
    \textbf{Velocity operator of am-Al$_2$O$_3$ for 120-atom AIMD model, and 120- and 7680-atom MACE models.}
    The square modulus of the velocity operator $\langle | \nu^{\mathrm{avg}}_{\omega_a\omega_d} |^2 \rangle$ is represented as  a function of energy differences $\hbar\omega_d = \hbar(\omega(\mathbf{q}_s) - \omega(\mathbf{q}_{s'}))$ and averages $\hbar\omega_a = \hbar\frac{\omega(\mathbf{q}_s) + \omega(\mathbf{q}_{s'})}{2}$, following Eq.~(\ref{eq:v_operator_omega_a_omega_d}). The upper panel shows a reasonable agreement between velocity operator elements of the 120-atom models generated with MACE (solid) and AIMD (dashed), both evaluated on a 5x5x5 $\bm{q}$-mesh. The middle panel shows that for MACE120, the velocity operator at $\bm{q}{=}0$ (dashed) is overall smaller than the velocity operator computed over a 5x5x5 $\bm{q}$-mesh (solid). The bottom panel shows that the velocity operator for MACE120 computed over a 5x5x5 $\bm{q}$-mesh (solid) is in reasonable agreement with the velocity operator of MACE7680 computed at $\bm{q}{=}0$ (dashed). 
    The delta functions used for calculation of $\langle | \nu^{\mathrm{avg}}_{\omega_a\omega_d} |^2 \rangle$ were replaced with Gaussian with variance related to the smearing parameter from the convergence plateau: $\sigma^2 = \eta^2 \pi / 2$. 
    }
    \label{fig:vel_op_mace_2D}
\end{figure}

\section{Allen-Feldman conductivity of MACE models}

To supplement the comparison of rWTE conductivities in Fig. \ref{fig:thermal_mace}, we present AF conductivities for AIMD and MACE models in Fig. \ref{fig:thermal_mace_AF}. We first note that for am-Al$_2$O$_3$ at densities close to 3.0 g/cm$^3$, influence of anharmonicity on the magnitude of conductivity is very weak, which is exemplified by very good agreement between WTE $\bm{q}{=}0$ and AF $\bm{q}{=}0$ predictions for 3240 and 7680-atom MACE models between 50 and 700 K. Our second conclusion is that predictions of AF conductivity using a $\bm{q}$-mesh in small models give the same trend and very similar magnitude as predictions of AF conductivity using $\bm{q}{=}0$ in large models for very disordered solids, which am-Al$_2$O$_3$ is an example of. The largest differences of value of AF conductivity at 700K are between MACE7680 at $\bm{q}{=}0$ and on 3x3x3 $\bm{q}$-mesh , and is equal to approximately $0.04$ W/mK ($2$ \%).\\[5mm]

\begin{figure}[H]
  \centering
\includegraphics[width=\WidthFigure]{\folder/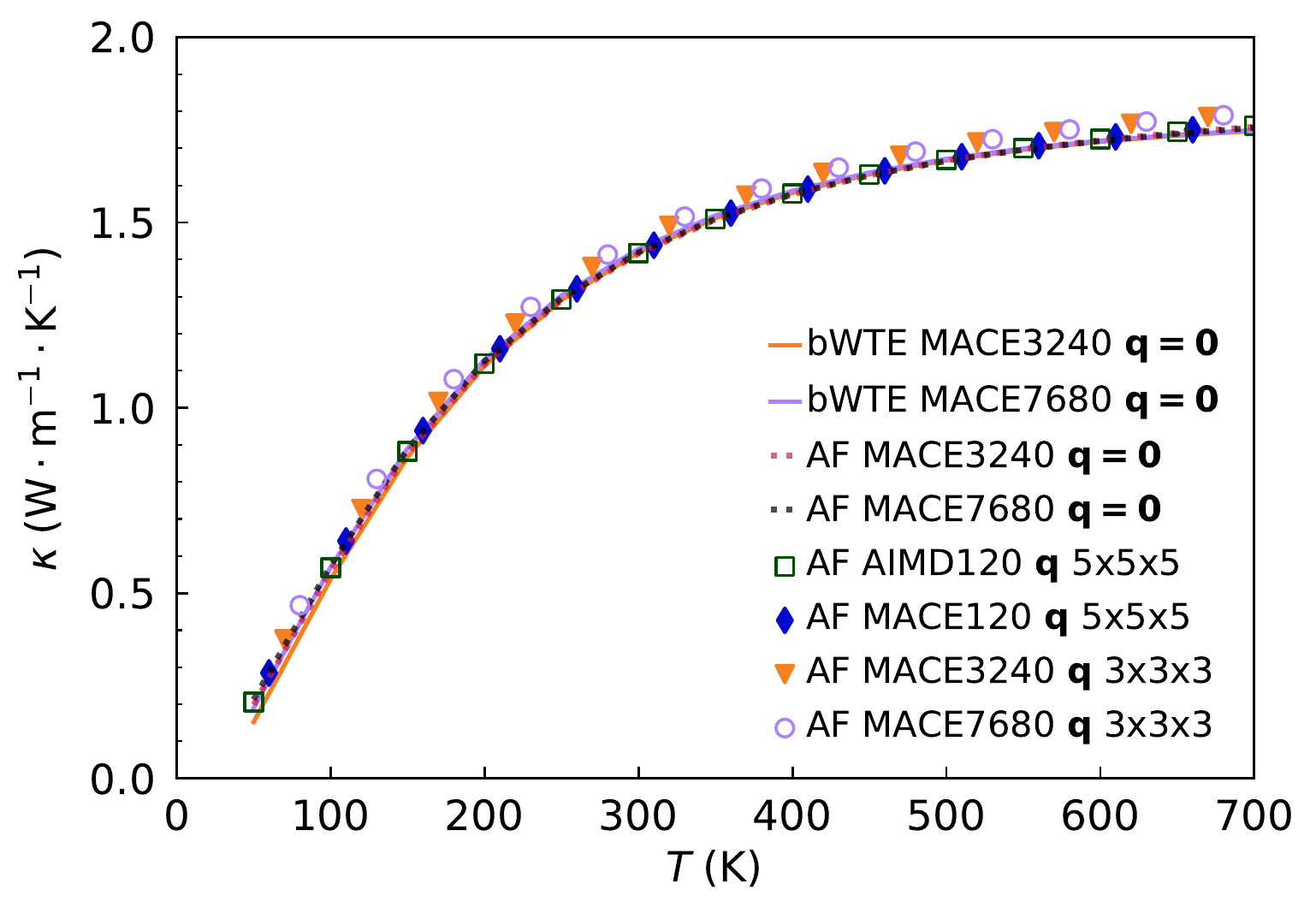}
  \caption{
  \textbf{AF thermal conductivity for models of am-Al$_2$O$_3$ with densities close to 3.0 g/cm$^3$.} Solid (dashed) lines are AF (WTE) calculations done at $\bm{q}{=}0$ for 3240-atom (orange) and 7680-atom (purple) MACE models. The scatter points are AF harmonic conductivities calculated on a mesh for 120-atom AIMD model (green, empty squares, 5x5x5 mesh), 120-atom MACE model (blue, filled diamonds, 5x5x5 mesh), 3240-atom MACE model (orange, inverted, filled triangles, 3x3x3 mesh) and 7680-atom MACE model (empty, purple circles, 3x3x3 mesh).}
  \label{fig:thermal_mace_AF}
\end{figure}

\providecommand{\noopsort}[1]{}\providecommand{\singleletter}[1]{#1}%

\end{document}